\documentclass[twoside,authoryear,preprint]{elsarticle}
\usepackage[T1]{fontenc}
\usepackage[latin9]{inputenc}
\usepackage{float}
\usepackage{rotfloat}
\usepackage{units}
\usepackage{textcomp}
\usepackage{amsmath}
\usepackage{amssymb}
\usepackage{graphicx}

\makeatletter

\providecommand{\tabularnewline}{\\}

\journal{Applied ocean research}

\@ifundefined{showcaptionsetup}{}{%
 \PassOptionsToPackage{caption=false}{subfig}}
\usepackage{subfig}
\makeatother

\usepackage[english]{babel}
\begin{document}

\title{Multivariate statistical modelling of future marine storms}

\author[LIM]{J.~Lin-Ye\fnref{fn1}\corref{cor1}}

\ead{jue.lin@upc.edu}

\author[LIM]{M.~Garc\'ia-Le\'on}

\author[LIM]{V.~Gr\`acia}

\author[ECA]{M.I. Ortego}

\author[CMCC]{P. Lionello}

\author[LIM]{A.~S\'anchez-Arcilla}

\cortext[cor1]{Corresponding author}

\address[LIM]{Laboratory of Maritime Engineering, Barcelona Tech, D1 Campus Nord,
Jordi Girona 1-3, 08034, Barcelona, Spain}

\address[ECA]{Department of Civil and Environmental Engineering, Barcelona Tech,
C2 Campus Nord, Jordi Girona 1-3, 08034, Barcelona, Spain}

\address[CMCC]{Centro Euro-Mediterraneo sui Cambiamenti Climatici, Via Augusto Imperatore,
16, Lecce, Italy}
\begin{abstract}
Extreme events, such as wave-storms, need to be characterized for coastal infrastructure design purposes. Such description should contain information on both the univariate behaviour and the joint-dependence of storm-variables. These two aspects have been here addressed through generalized Pareto distributions and hierarchical Archimedean copulas. A non-stationary model has been used to highlight the relationship between these extreme events and non-stationary climate. It has been applied to a Representative Concentration pathway 8.5 Climate-Change scenario, for a fetch-limited environment (Catalan Coast). In the non-stationary model, all considered variables decrease in time, except for storm-duration at the northern part of the Catalan Coast. The joint distribution of storm variables presents cyclical fluctuations, with a stronger influence of climate dynamics than of climate itself.\end{abstract}
\maketitle

\section{Introduction\label{sec:Introduction}}

Extreme events characterization is a key piece of information for an efficient design and construction of any
coastal infrastructure. Natural extreme events, such as hurricanes,
tsunamis or earthquakes, can lead to considerable economic losses
\citep{Shi2016}. From all these hazards, marine storms cause most of the damage to non-seismic coasts. This situation may eventually
be aggravated as a consequence of Climate-Change, which affects the
intensity and frequency of extreme wave-conditions \citep{Wang15a,Hemer2016a}. 

Changes in climate can affect several coastal hazards: flooding \citep{Hinkel14PNAS,wahl16a},
erosion \citep{Hinkel13,Casas-Prat2016,Li201453}, harbour agitation
\citep{ASA2016a,Sierra15} and overtopping \citep{Sierra2016}. A robust statistical
characterization of storms is, thus, required to assess coastal risks and to forecast storm impacts \citep{ASA14,Gracia13}. The stationary climate assumption,
common approach in the last decades for designing infrastructures,
does no longer hold valid in a context of Climate-Change. Hence, there is
a pressing urge for methodologies that consider non-stationarity,
not only in trends, but also in higher statistical moments such as variance. 

Usual statistical distributions for extremes such as the Generalized
Pareto Distribution (GPD) or the Generalized Extreme Value distribution 
have three parameters: location, scale and shape. \citet{Rigby05}
proposed a generalized additive model for these three parameters to
predict river flow-data from temperature and precipitation on the
Vatnsdalsa river (Iceland). \citet{Yee07} developed a methodology
that allows extreme value distributions to be modelled as linear or
smooth functions of covariates. One of the examples they presented was
the modelling of rainfall in Southwest England. 
\citet{Du2015} carried out frequency analyses using meteorological
variables, where they tested several combinations of co-variates with generalized
additive models for location, scale and shape, and concluded that meteorological
co-variates improve the characterization of non-stationary return
periods. \citet{mendez2007analyzing} used a time-dependent 
generalized extreme value distribution to fit monthly maxima series of a large historical tidal gauge record,
allowing for the identification and estimation 
of time scale such as seasonality and interdecadal variability. 
\citet{mendez2008seasonality} extended the former methodology to significant wave-height, 
while considering the effect of storm duration. 

For design purposes, the most analysed variable in marine storms
is the significant wave height ($H_{s}$), usually considered
to be independent from other wave storm-components such as peak-period
($T_{p}$), or storm-duration ($D$). Nevertheless, these variables
are known to be semi-dependent \citep{DeMichele2007}. Univariate
analyses on singular variables, such as $H_{s}$, cannot thus describe
coastal processes adequately \citep{Salvadori2014}, leading
to misestimation of coastal impacts and risks.

The relationship among storm variables can be modelled with
statistical techniques such as parametric probability distributions
\citep{Ferreira200231}, asymptotic theory \citep{Zachary1998}, joint
modelling \citep{BitnerGregersen2015279}, or copulas \citep{Genest2007,Trivedi2007},
among other techniques. Copulas were proposed by \citet{Sklar_seminal_1959},
and have recently attracted attention from coastal engineers \citep{Corbella2012,Salvadori2015}.
\citet{wahl2011multivariate} applied fully nested Archimedean copulas
to wave storms off the German coast. They first characterized the
highest energy point and its intensity and then incorporated the
significant wave height. Complementary to these methodologies, \citet{Gomez16}
has implemented a time varying copula to analyse the relationship
between air temperature and glacier discharge, which is non-constant
and non-linear through time. In this case, both marginal and copula
parameters depend on time, and a full Bayesian inference has been applied
to obtain these parameters. 

Based on this, the present work characterizes
the extreme wave climate under a Representative Concentration Pathway 8.5 Climate-Change scenario  
(RCP8.5, i.e. an increase of the radiative forcing values by year 2100 relative
to pre-industrial values of $8.5\mathrm{W/m^{2}}$; \citet{IPCC2013}) for
a fetch-limited environment (Catalan coast). The study is based on a set of geographical nodes which are equidistant along the Catalan coast. Only eleven nodes out of the total twenty-three are used in this paper, since they represent well the main features and spatial variability of the storm distributions (see Fig. \ref{fig:Study-area}, red triangles). 
Two of the eleven nodes are in intermediate waters, while the rest are in deep waters. The subsequent analysis is performed assuming, first, stationary, and then, transient conditions. 

Section \ref{sec:Methods} describes the methodology and the theoretical
background. Section \ref{sec:Study-area} presents the study area.
Section \ref{sec:Results} lists main results, which are discussed in Section \ref{sec:Discussion}. The conclusions are summarized in Section \ref{sec:Conclusions}. 

\section{Study area\label{sec:Study-area}}

The Mediterranean Sea (see Fig. \ref{fig:Study-area}) is a semienclosed
basin, constrained by the European, Asian and African continents.
It has a narrow connection to the Atlantic Ocean (Gibraltar Strait),
as well as an access to the Black Sea. In terms of waves, the Mediterranean
Sea can be splitted into different partitions \citep{Lionello05}.
This paper deals with the Catalan coast, which can be found at
the northwestern Mediterranean sector. This area has, as its main morphological
features, a) mountain chains which run parallel and adjacent to the coast,
b) Pyrenees Mountains to the north, and c) the Ebre river valley to the
south. These orographic discontinuities, along with the major river
valleys, serve as channels for the strong winds that flow towards
the coast \citep{Grifoll2015}.

The most frequent and intense wind in the Catalan Coast is the Tramuntana
(north), appearing in cold seasons. It is the major forcing for the northern
and central Catalan Coast waves. However, from latitude $41^{\circ}N$ southward,
the principal wind direction is the Mistral (northwest), which is formed
by the winds that flow downhill the Pirinees or between the
gaps of the mentioned mountains. A secondary wind, the Ponent (west), comes
from the depressions in northern Europe. It is the second most frequent one, with limited intensity. Eastern winds are the ones with larger fetch for intense sheer stress, corresponding to low pressure centres over the northwestern Mediterranean.
During the summer, there are southern sea-breezes and estern winds, triggered by an intense high-pressure area on the British Islands. 

The northwestern Mediterranean Sea is a fetch-limited environment, primarily
driven by wind-sea waves \citep{Bolanos09,ASA2016a}. The distance
that waves travel, from the storm genesis to the Catalan Coast, is
at most one-sixth that of a wave that reaches the Atlantic European
coasts \citep{G93}. Therefore, the corresponding wave-periods, in
the northwestern Mediterranean, are much shorter. 

The present climate presents a mean significant wave height $\overline{H_{s}}$
of $0.72\unit{m}$ from Barcelona City nortward, and $0.78\unit{m}$ southward. Maximum
$H_{s}$ ranges between $5.48\unit{m}$ in the southern coast to $5.85\unit{m}$
at the northern coast \citep{ASA08,Bolanos09}. \citet{Casas-Prat2013} projected future wave climate
at the Catalan Coast through Regional Circulation Model outputs from the A1B scenario \citep{IPCC2000}
for the time-period comprising 2071-2100. Their results
showed a variation compared to present of the significant
wave height around $\pm10\%$, whereas the same variable for a $50\unit{year}$
return-period exhibits rates around $\pm20\%$. 

\section{Proposed methodology\label{sec:Methods}}

The methodology here developed leads to a robust assessment of storm pressures under present or future climates. Regional projections are obtained from a deterministic approach, based on the underlying physics, avoiding the computationally expensive dynamical downscaling and the oversimplification of conventional empirical downscaling. Wave storms are first characterized assuming stationarity (see Fig. \ref{fig:Flow-chart}). From here, the joint probability structure is derived and this will serve as a basis for the non-stationary model of the selected projection (in this case, under the RCP 8.5 scenario). A non-stationary model is then built, and constitutes the main part of the proposed methodology, described below.

\subsection{Data and storm components}

The analysis has been performed considering the wave-climate at
the Catalan Coast under a RCP 8.5 Climate-Change scenario.
This scenario considers a $CO_{2}$ concentration in the atmosphere
close to $1250\unit{ppm}$ in 2100, which is double that of any other
scenario in the Fifth Assessment Report \citep{IPCC2013}. The modelling
chain comprises the CMCC-CM \citep{Scoccimarro11} 
Global Circulation Model (see Table \ref{tab:GCM_models}), providing boundary conditions for
the Regional Circulation Model COSMO-CLM \citep{Rockel08}. The statistical model derived from the CMCC-CM dynamical downscaling has been validated with a total of eighteen Global Circulation Models,
shown in Table \ref{tab:GCM_models}. This list includes models
from the same experiment (CMIP5, \citet{Taylor12a}) and from the same Climate-Change-scenario
(RCP 8.5), covering, thus, a comprehensive range of predictors. 
The COSMO-CLM grid, that has a resolution of $0.125^{\circ}\times0.125^{\circ}$,
spans the whole Mediterranean region. The next step consists of the WAM \citep{WAMDI88} wave model, where the just mentioned wind fields serve as an input, for the same domain
and spatial resolution. The projections considered in all three models (Global Circulation Model, Regional Circulation
Model and WAM), span the interval from year 1950 to 2100. 

The nodes considered for the AR5 projections and subsequent analyses (Fig. \ref{fig:Study-area}, red triangles) are combined with buoy and SIMAR \citep{Gomez2005} hindcast points (green rhombuses and black dots, respectively) for validation purposes. All selected nodes (except 1 and 16) are located in deep waters, and thus the WAM model is a suitable option \citep{Larsen15b}. The application of this code to nodes 1 and 16, in intermediate waters, may present certain limitations and would, thus, require further exploration and research. The validation dataset comes from SIMAR hindcasts and Puertos-del-Estado buoy records, corresponding to the period 1990 to 2014. Storms here are clustered into storm-years. Storm-years (called ``years'',
hereafter), which are periods of 12 months, from $1^{st}$ July to $30^{th}$ June of the next year.  

Four main variables have been selected to describe the storm-intensity conditions: storm energy
($E$), significant wave-height at the storm-peak ($H_{p}$), peak wave-period at the storm-peak
($T_{p}$), and duration ($D$). The $E$ and $D$ are aggregated parameters, related to the total impact of the storm, whereas $H_{p}$ and $T_{p}$ represent the maximum intensity of the event. $E$, $H_{p}$, $T_{p}$ and $D$ take positive
real values and, consequently, they have been log-transformed to avoid scale
effects \citep{egozcue2006effect}.

\subsection{Pre-analysis (stationarity assumption)\label{sub:Pre-analysis}}

Prior to the actual modelling, an explanatory analysis has been carried out
with the available wave data. A set of stationary models has been built by selecting
equidistant time slices from the total sample, following previous work by other authors
with similar hydrodynamic variables \citep{muis2016nat,Vousdouskas2016}.
The three time-frames are labelled as: (i) past (PT,1950-2000); (ii)
present-near-future (PRNF, 2001-2050), and far future (FF, 2051-2100).
Storms have been defined using a stationary $H_{s}$ threshold of $2.09m$ significant wave-height, 
based on previous work \citep{LinYe16}. Although the time period in 
\citet{LinYe16} is significantly shorter than in the present paper, this threshold should be
acceptable for the three time-frames as it falls on the linear part of the
excess-over-threshold plot (Fig. \ref{fig:excess-over-threshold}), according 
to methodology previously developed by \citet{Tolosana-Delgado2010}. 

The next step of the pre-analysis consisted in building dependograms of the selected storm variables, which were then  visually inspected for non-stationary behaviour. Each variable is also presented in absolute concentration curves (ACC),
where ACC1 indicates the ratio of $q_{50}$ at a given time-frame,
to the one in the PT inteval \citep{yitzhaki1991concentration}. ACC2 denotes the same ratio, but with $\left(q_{75}-q_{50}\right)$. 
Thus, ACC1 represents on changes in the mean, whereas ACC2 reflects on the evolution of the variance.
This analysis has been performed for the energy and duration of the total events of a storm-year, $E_{year}$ and $D_{year}$,
as well as the mean $H_{s}$ and $T_{p}$ of a storm-year, $\overline H_{s,year}$ and $\overline T_{p,year}$,
to assess non-stationary trends.

\subsection{Stationary model\label{sub:Methods-Stationary-model}}

The probability distribution of each storm variable is fit
by a GPD. Being $Y=X-x_{0}$ the excess of a magnitude $X$ over a
location-parameter $x_{0}$, conditioned to $X>x_{0}$, the support
of $Y$ is $\left[0\,,\, y_{sup}\right]$ \citep{Coles2001}. $y_{sup}$ is the upper bound of the GPD. The
GPD cumulative function is, then,
\begin{equation}
F_{Y}\left(y|\beta,\xi\right)=1-\left(1+\frac{\xi}{\beta}y\right)^{-\frac{1}{\xi}}\ ,\ 0\leq y\leq y_{sup},
\end{equation}
where $\beta\geq0$ is the scale parameter and $\xi\in\mathbb{R}$
is the shape parameter. 
As a first approximation, the values of the location parameters $x_{0}$ 
obtained in \citet{LinYe16} have also been used in this case. The 
departure from these values is described in 
Sub-section \ref{sub:Results-Stationary-model}.

The Hierarchical Archimedean copula (HAC) is a flexible tool that
describes the dependence between variables via the nesting of a subset of
$2$-D copulas \citep{Sklar_seminal_1959,nelsen2007introduction,Okhrin2013}.
The Gumbel type HAC with a mean aggregation method is selected for this
case of extreme events, according
to \citet{LinYe16}. A $d$-dimensional Archimedean copula has the
form 
\begin{equation}
\mathrm{C}\left(\mathbf{F};\text{\ensuremath{\phi}}\right)=\mathrm{\phi^{-1}}\left(\text{\ensuremath{\phi}}\left(F_{1}\right)+\cdots+\phi\left(F_{d}\right)\right),\quad\mathbf{F}\in\left[0,1\right]^{d},
\end{equation}
for a given generator function $\mathbf{\phi}$. A Gumbel generator
has been selected since it defines the dependence in the upper tail of
the probability distribution. Note that a family of asymmetric copulas
\citep{vanem2016joint} would include physical limitations, such as
wave steepness, where high $H_{p}$ cannot commute with large $T_{p}$.
Due to the complexity of non-stationarity, the asymmetric copulas 
must be carefully introduced in a more mature future version of the proposed
model. 

The HAC aggregates the Gumbel generator parameters using a series of
coefficients called $\theta$, which can be transformed to Kendall's
$\tau$ \citep{Kendall37,Salvadori2011}. $\tau$ denotes independence
when $\tau=0$, and total dependence when $\tau$ tends to $1$.
The goodness-of-fit of the HACs at each time-frame has been
assessed by using goodness-of-fit plots of the empirical copulas \citep{LinYe16}.
The $\kappa^{2}$ statistic (\cite{Gan1991}) serves to quantify the goodness-of-fit.
It takes values in $\left[0,1\right]$, and a perfect fit happens
when $\kappa^{2}=1$.
According to our experience in the Catalan Coast, the HAC-structure in Fig. \ref{fig:HAC-tree}
should be applicable to this area. There is another approach for events where $H_{p}$ 
is less inter-dependent with $E$ and $D$ \citep{LinYe16},
but this type of structure is of less interest in this study, as will be discussed later. 
The nesting levels in Fig. \ref{fig:HAC-tree} start at the branching of the tree-like structure, and
end at the top "root" level.

\subsection{Non-stationary model\label{sub:Methods-Non-stationary-model}}

Extreme events are scarce by nature. The shorter the time-window considered,
the smaller will be the available information, with larger uncertainty.
This assumption means that, for the time-windows of $50\unit{years}$
considered in the stationary model, there are fewer samples of high
extreme events. Hence, the probability distribution function's upper
tail estimation would not provide results reliable enough. Previous
studies indicate that Climate-Change also has a non-negligible effect
on extremes \citep{Trenberth15,Hemer2016a,Du2015}, so assumptions
such as a stationary storm-threshold cannot be adopted. This is a
first indication that non-stationarity needs to be addressed \citep{Vanem2015}.

In the non-stationary model, vectorial generalized additive models
(VGAM, \citet{Yee96a}) have been used to determine storminess,
storm-thresholds and GPD parameters \citep{Rigby05,Yee07}.
The VGAM consists of a linear function \citep{fessler1991nonparametric,hastie1990generalized}:
\begin{equation}
\eta_{i\left(j\right)}=\beta_{1\left(j\right)}^{*}+\mathrm{f_{2\left(j\right)}}\left(x_{i2}\right)+\ldots+\mathrm{f_{p\left(j\right)}}\left(x_{ip}\right),
\end{equation}

\noindent where $\eta_{i\left(j\right)}$ is the $j^{\mathrm{th}}$ dependent
variable, $x_{i}$ is the $i^{\mathrm{th}}$ independent variable that generates
$\eta_{i}$. $\eta_{i}$ is a sum of smooth functions of the individual
covariates $\beta_{1\left(j\right)}^{*}$ and $f_{p\left(j\right)}$.
In this case, $\beta^{*}$ is not the scale parameter of the GPD. Additive
models do all the smoothing in $\mathbb{R}$, avoiding the large bias
introduced in defining areas in $\mathbb{R}^{n}$. 

The mathematical assumptions for regression models are: 1) incorrelation, 
2) normality, and 3) homoscedasticity of residuals. Assumption 1) is assessed with a ACF plot,
assumption 2) can be assessed with a Q-Q plot against a $N\left(0,\sigma^{2}\right)$
distribution, where the sample standard deviation is used as $\sigma^{2}$.
Assumption 3) can be analysed on a graph of fitted value vs. residuals.
When the predicted variable is a counting one, a vectorial
generalized linear model (VGLM) can be adopted \citep{Yee96a}. The VGLM
is a particular case of VGAM. The storminess is a counting variable,
and its relationship with any other factor can be approximated by a Poisson
distribution.

The storm-threshold is then estimated through a VGAM that approximates
its relationship with a factor by a Laplace distribution. Once storms
are selected, their non-stationary GPD location-parameter $x_{0}$
is estimated through quantile regression \citep{Koenker05a}. The
quantile regression is a specific type of VGAM, and it estimates the
$100\hat{\tau}\%$ conditional quantile $y_{\hat{\tau}}\left(x\right)$
of a response variable $Y$ as a function $u\left(x,\tau\right)$
of covariates $x$. The equation $l_{u}^{*}=l_{u}+\varrho_{u}R_{u}$
must then be minimized, where $l_{u}=\hat{\tau}\underset{i:r_{i}\geq0}{\sum}\left|r_{i}\right|\left(1-\hat{\tau}\right)\underset{i:r_{i}<0}{\sum}\left|r_{i}\right|$
for residuals $r_{i}=y_{i}-u\left(x_{i},\hat{\tau}\right)$. $\varrho$
is a roughness coefficient that controls the trade-off between quality
of fit to the data and roughness of the regression function; and $R$
is a roughness penalty \citep{Northrop11a,Jonathan13a}. The above mentioned
$\hat{\tau}$ has nothing to do with the $\tau$ of Kendall. Regarding
the rest of the GPD parameters: $\xi$ is assumed to remain constant; 
$\beta$ is considered to depend on co-variates, and is estimated
with VGLMs. 

The option of using time as a covariate is examined in the non-stationary
model, just to assess the evolution of other variables. 
The predicting function is a $4$-degree spline \citep{hastie1990generalized}.
Alternative predictive parameters seems to
present a greater potential. Climate-indices are eligible
candidates \citep{Rigby05}, for which the linear interpolation function has been selected, advocating the principle of parsimony. Possible climate-indices are the North Atlantic Oscillation
(NAO, \citet{Hurrell2009}), the Easterly Atlantic index (EA, \citet{Barnston87a}),
the Scandinavian oscillation (SC, \citet{Barnston87a}), and their first
and second time derivatives. These climate-indices have been scaled to have a mean
value equal to zero and a variance equal to unity, and they actually introduce
time as an implicit covariate. They were computed from the monthly-averaged
sea level pressure fields, from the global circulation-model listed in Table \ref{tab:GCM_models}.
In order to avoid sudden oscillations that would hinder
interpretation, the time series of climate-indices have been filtered with
a $2^{nd}$ order lowpass Butterworth filter \citep{butterworth30},
whose low-pass period was of $10\unit{years}$.

Different results among global circulation-models should be expected, despite the same
post-processing treatment for all of them. 
The grid-size and physical implementations are not the same,
the model with the highest resolution ($0.76{}^{\circ}\times0.76{}^{\circ}$)
is CMCC-CM, which is the one that has served as the
calibration model. There are also slight divergences on how the model
addresses the evolution of emissions \citep{Friedlingstein14}.

\begin{figure}[tph]
\noindent \begin{centering}
\includegraphics[bb=50bp 150bp 662bp 550bp,clip,width=15cm]{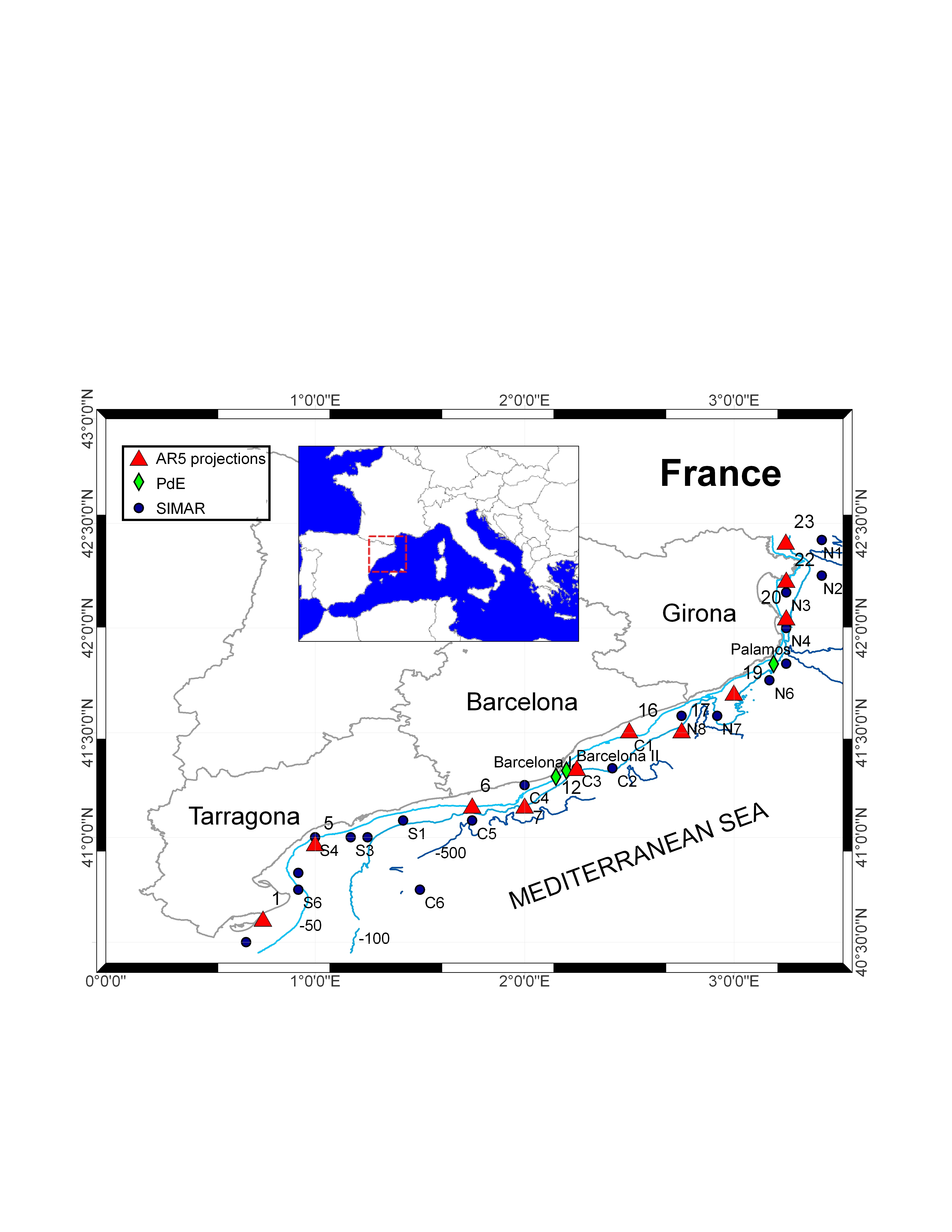}
\par\end{centering}

\caption{Map of the Catalan Coast, area located in the northwestern Mediterranean.
The bathymetry is in meters, showing how all nodes where
the proposed model applies (AR5 nodes) are in deep water, except nodes 1 and 16. AR5 nodes are represented by 
red triangles, buoy (PdE) nodes are green rhombuses, and SIMAR nodes
are solid black points.\label{fig:Study-area}}
\end{figure}

\begin{figure}[H]
\begin{centering}
\includegraphics[bb=0bp 60bp 540bp 680bp,clip,height=14cm]{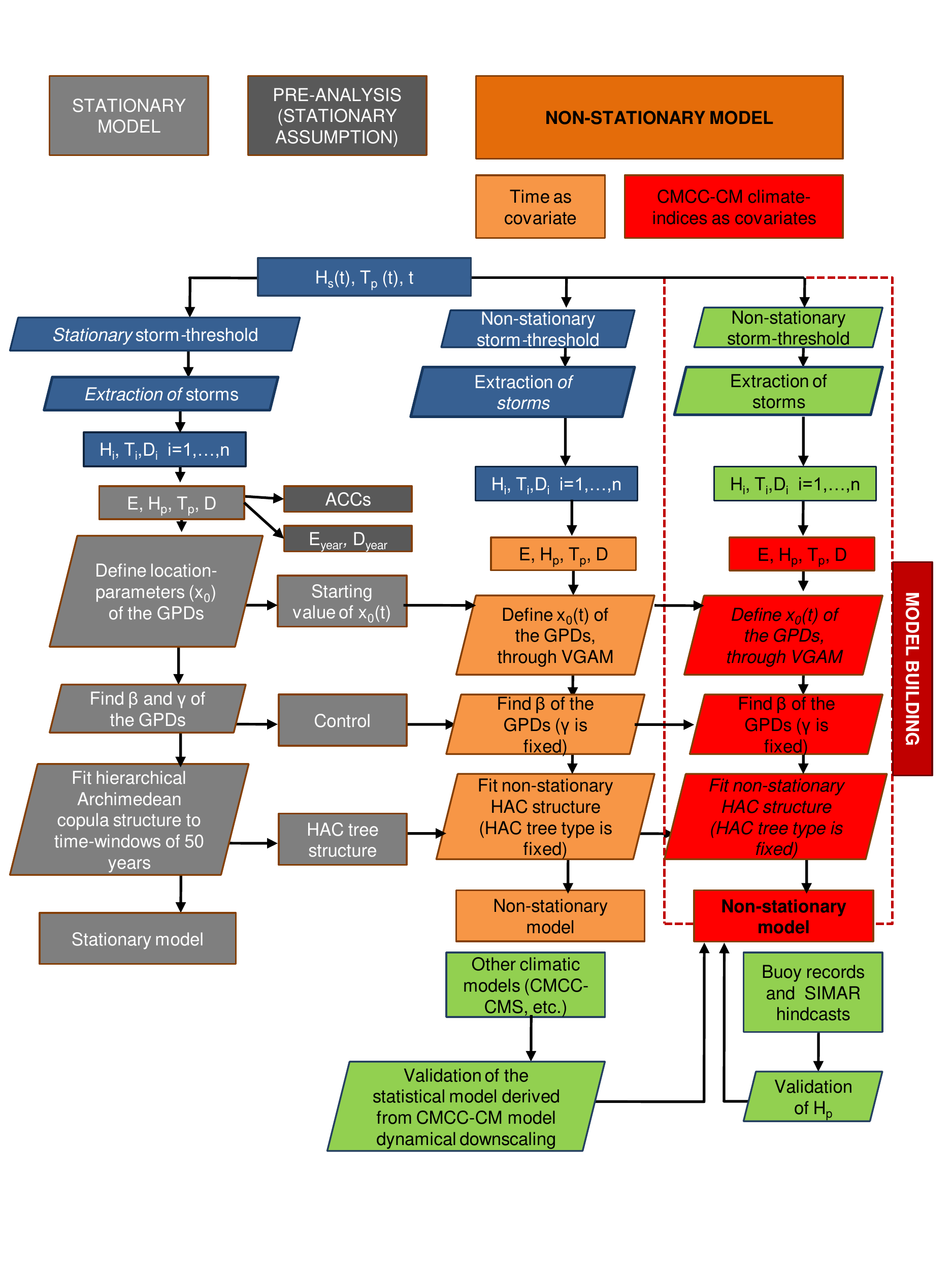}
\par\end{centering}

\caption{Flow-chart of the methodology applied in this paper.\label{fig:Flow-chart}}
\end{figure}

Once storms events have been selected, $E$, $D$, $H_{p}$ and $T_{p}$ can be extracted.
The effect of climate-indices as covariates is assessed at nodes
7 and 21, as these nodes represent the most distinct spatial patterns
(see Sec. \ref{sec:Study-area} and Fig. \ref{fig:Study-area}). The
goodness-of-fit of the resulting VGAM with different combinations
of covariates is contrasted with a likelihood-ratio test (LRT, \citet{Vuong89}),
the Akaike information criterion (AIC, \citet{akaike1987factor})
and the Bayesian information criterion (BIC, \citet{tamura1991procedure}). 
A censorship analysis is carried out on the sample for these two nodes, corresponding to two subsets of GPDs for: a) onshore winds and b) offshore winds. 
For the two samples in the censorship analysis, and for the combined sample, the proposed model is calibrated with climate-indices derived from the CMCC-CM global circulation-model. The climate-indices from the other eighteen models 
(Figs. \ref{fig:NAO}, \ref{fig:EA}, and \ref{fig:SC}) serve to
predict what would be the probability distribution functions under
a wide range of plausible values.
In the results and discussion section, the $99^{\mathrm{th}}$ quantile, a common quantile for hazard and design \citep{Goda10}, has been
used to inter compare these.

VGAM uses, thus, global circulation climate-indices as covariates to create time series
of $99^{\mathrm{th}}$ quantiles. A way of quantifying how these time series
differ from the baseline (CMCC-CM), is by computing the Euclidean
distance between the estimated partial autocorrelation coefficients
of each time series (\citet{Galeano00}). This metric takes values in
$\left[0,1\right]\in\mathbb{R}$, being $0$ the shortest distance
(i.e. closer similarity between models), and $1$, the largest one.

Regarding the joint dependence structure of the proposed model, 
storms are clustered into periods of $15\unit{years}$,
under the assumption that there is stationarity in these $15\unit{years}$.
Because of the persistence of the climate-indices considered,
this is a plausible hypothesis. $15\unit{years}$ are also
the shortest time-span that provides a sufficient number of storms
to determine the HAC structure. Larger time-windows would offer a
greater number of storms, but with a non-stationary dependence parameter. Non-stationary HAC dependence parameters are obtained at each node, for this moving time-window of $15\unit{years}$.
Each time-window overlaps with the former and the following ones,
in half-a-year, to characterize the non-stationary effect. 

The Gumbel HAC dependence structure from the stationary-model is also used in the non-stationary model. 
Particularly, the HAC-structure in Fig. \ref{fig:HAC-tree} is adopted for the whole non-stationary model.
The fitting criteria
is the Maximum Likelihood method, where the HAC-structure in the stationary-model
(see sub-section \ref{sub:Methods-Stationary-model}) is set as the
unique structure for all nodes and for the whole simulation period.
The selection of only one HAC-structure follows the principle of parsimony,
being this HAC the one that better characterizes the joint-dependence
at most spatial nodes during the three time-frames of the stationary model.

The Kwiatkowski-Phillips-Schmidt-Shin (KPSS) test \citep{KWIATKOWSKI1992159}
is applied to the dependence-parameters of the HAC, to look into the
stationarity of the $\tau$ time series. The p-value of
such test gives the level of significance at which the null test cannot
be rejected. In other words, on how likely the dependence-parameter
is actually stationary.

To represent projected climatology,
the probability distribution function of the $H_{p}$ should resemble
that of observed storm conditions (from buoys and hindcasts). The proposed model has been validated at the nodes listed on Table \ref{tab:validation}
(see Figs. \ref{fig:Study-area} for node location), as follows. The SIMAR/buoy
data validation nodes are denoted: 
\begin{equation}
\left\{ H_{p,1},\ldots,\mbox{\ensuremath{H_{p,i}}},\ldots,H_{p,n}\right\} ,\quad i=1\div n,\ n\in\mathbb{R},
\end{equation}
and the model data (written as $H_{p}^{*}$, here)
\begin{equation}
\left\{ H_{p,1}^{*},\ldots,\mbox{\ensuremath{H_{p,j}^{*}}},\ldots,H_{p,m}^{*}\right\} ,\quad j=1\div n,\ m\in\mathbb{R}
\end{equation}
They are next combined to form a joint dataset: 
\[
\left\{ H_{p,1},\ldots,\mbox{\ensuremath{H_{p,i}}},\ldots,H_{p,n},H_{p,1}^{*},\ldots,\mbox{\ensuremath{H_{p,j}^{*}}},\ldots,H_{p,m}^{*}\right\} 
\]

Such set is partitioned into four intervals, separated by
the quartiles $\qquad$ $\left\{ q_{0},q_{25},q_{50},q_{75},
q_{100}\right\} $. There are elements from both SIMAR/buoy $H_{p}$ and AR5 projections,
in each interval. The quartiles are selected as boundaries
because buoy records are often interrupted due to harsh wave conditions.
Then, if the selected intervals are too small, some of them might
be empty, which would lead to indetermination of the distance
between model and data. 

Two vectors are defined as 
\begin{equation}
vec_{obs}=\left(\sum_{q_{0}}^{q_{25}}p\left(H_{p,i}\right),\sum_{q_{25}}^{q_{50}}p\left(H_{p,i}\right),\sum_{q_{50}}^{q_{75}}p\left(H_{p,i}\right),\sum_{q_{75}}^{q_{100}}p\left(H_{p,i}\right)\right),\label{eq:aitch-obs}
\end{equation}

and

\begin{equation}
vec_{model}=\left(\sum_{q_{0}}^{q_{25}}p\left(H_{p,j}^{*}\right),\sum_{q_{25}}^{q_{50}}p\left(H_{p,j}^{*}\right),\sum_{q_{50}}^{q_{75}}p\left(H_{p,j}^{*}\right),\sum_{q_{75}}^{q_{100}}p\left(H_{p,j}^{*}\right)\right),\label{eq:aitch-mod}
\end{equation}
where $vec_{obs}$ is the vector for observations, and $vec_{model}$
is the one for projections. Each element of the vector is the summation
between two quantiles of the probability distribution function. Therefore, $vec_{obs}$
and $vec_{model}$ are compositional data, their elements being parts
of a whole \citep{egozcue2011evidence}, and fulfilling some other
properties defined in \citet{aitchison1982statistical} and \citet{egozcue2003isometric}.
The distance between these two vectors can be determined with an Aitchison
measure \citep{aitchison1992criteria,pawlowsky2001geometric},
\begin{equation}
d\left(\mathbf{x},\mathbf{y}\right)=\left|\ln\frac{\mathbf{x}\left(\mathbf{1}-\mathbf{y}\right)}{\mathbf{y}\left(\mathbf{1}-\mathbf{y}\right)}\right|,\quad\mathbf{x},\mathbf{y}\in\left(0,1\right)\in\mathbb{R},\label{eq:Aitchison-distance}
\end{equation}

Where $\mathbf{x}$ and $\mathbf{y}$ are two compared vectors. Another
measure for the distance is the Kullback-Leibler divergence \citep{kullback1997information}
\begin{equation}
D_{KL}\left(P\parallel Q\right)=\underset{i}{\sum}P\left(i\right)\log\frac{P\left(i\right)}{Q\left(i\right)}.\label{eq:KL}
\end{equation}
This function measures the extra entropy of the probability distribution
$Q$ of the model, with respect to the probability distribution $P$ of
the observations. Note that for any $i$, $Q\left(i\right)=0$, must
imply $P\left(i\right)=0$, to avoid indertemination, thus ensuring that the model considers all the values that the observations
show. Also, whenever $P\left(i\right)=0$, the contribution of
the $i$-th term is null, as $\underset{x\rightarrow0}{\lim}x\log\left(x\right)=0$.

Both eq. \ref{eq:Aitchison-distance} and \ref{eq:KL} are distances,
and thus take values in $\mathbb{R}_{0}^{+}$. The module of the vector
is a particular case of both distances \citep{egozcue2011evidence}, and thus both can be compared
to the vectorial module, in Euclidean space, of $\mathbf{x}$ and
$\mathbf{y}$, which should be of order $1$.

\section{Results \label{sec:Results}}

\subsection{Pre-analysis (stationarity assumption)}

The dependograms, which do not vary for the different time-frames, show inter-dependence of $T_{p}$ and
the other variables ($E$, $H_{p}$, $D$), except at node 1 in the
FF.  
ACC1 and ACC2 ratios are represented in Figs. 1 to 3 
of the Supplementary material.
$E$ and $D$ decrease in PRNF and FF (see Supplementary material, Fig. 1). 
$ACC1_{H,prnf}$, $ACC1_{H,ff}$, $ACC1_{T,prnf}$ and $ACC1_{T,ff}$ are equal to one for the entire
Catalan Coast (figures not shown). 
$ACC1_{E,prnf}$ is slightly below 
1, being specially low in bays or similar local coastal domains.
$ACC1_{E,ff}$ is approximately 1.05 in the northern sector (Girona). $ACC1_{D,prnf}$
and $ACC1_{D,ff}$ are high in apexes like the Creus cape (near node 22), and low
in bays like the Tarragona one (see Fig. \ref{fig:Study-area}). All the ACC2 ratios are slightly below 
one in the PRNF (see Supplementary material, Fig. 2), 
 and get closer to one in the FF (see Supplementary material, Fig 3). 
 The temporal evolution of $E_{year}$, $\overline H_{s,year}$, $\overline T_{p,year}$
and $D_{year}$ are presented in Figs. 4 to 7 
of the Supplementary material.
The $E_{year}$ are only autocorrelated
at node 22 and 12, with a lag of $9\unit{years}$ in PT, and are not
autocorrelated for larger lags. $\overline H_{s,year}$ is 
autocorrelated at nodes 6, 12, 16, 17, 20, 22 and 23, at different time-frames, 
and $\overline T_{p,year}$ is autocorrelated along the entire Catalan coast.  
$D_{year}$ is autocorrelated at node
22, in PT, with a lag of $5\unit{years}$, and at node 1 in PRNF, with
a lag of $2\unit{years}$.

\subsection{Stationary model\label{sub:Results-Stationary-model}}

After defining the GPD parameters $x_{0}$ and $\beta$, each storm-intensity
variable is fit by a GPD, of discontinuous support. 
$T_{p}$ has required an increase of its location-parameter
($10\%$ in FF, at nodes 20 and 22), before fitting GPD. Depending on 
location, differences may appear within storm-parameters, possibly
due to wave propagation effects and the control of land
winds at the northermost and southernmost sectors. Unlike for SIMAR hindcasts,
the HAC-structure in Fig. \ref{fig:HAC-tree}
is the only one present at all nodes and for all time-frames. The goodness-of-fit of the
HAC are represented in Figs. 8 to 10
of the Supplementary material. 
The $k^{2}$ parameter and the
graph show a good fit of the Gumbel-HAC, as observed in \citet{LinYe16}.

\subsection{Non-stationary model\label{sub:Non-stationary-model}}

Two different kinds of non-stationary model have been built: a) using
time as the single covariate (NS-T hereafter); and b) implementing large scale climate-indices as covariates (NS-CI hereafter).
By using time alone as a covariate to storminess,
the storm threshold and GPD parameters, whenever NS-T shows a clear time-dependent behaviour, the non-stationary model NS-CI is applicable.
Figures \ref{fig:tau-time-root}, \ref{fig:tau-time-EDH}, and \ref{fig:tau-time-ED}
show the temporal evolution of the HAC dependence-parameters for
NS-T. The KPSS test \citep{KWIATKOWSKI1992159} is applied on 
$\tau$ for the NS-T model, and the outcome is that the null hypothesis
of stationarity cannot be rejected in $1-4\%$ of the cases. That
is, $\tau$ is highly non-stationary.

Regarding storminess, the SIMAR-dataset and the available buoy-records confirm
higher storminess-indices ($\lambda$) at the northern coast (Figs.
\ref{fig:lambda-time} and \ref{fig:lambda-NAO}). Figure
\ref{fig:lambda-time} shows
that $\lambda$ decreases with time, but the stationary model can
only capture this trend via the predefined time-blocks. This
supports using a non-stationary model to improve the
representation of the extreme wave-climate. A sensitivity analysis has been carried out on the covariates, at nodes 7 and 21. 
In the censorship analysis within this sensitivity analysis, 
the subset with on-shore winds has presented better fit with NAO as covariate, whereas the subset with 
offshore-winds has done the same with SC. However, an additional test on the rest of nodes has not shown better performance, and for the sake
of consistency and parsimony, the uncensored sample has been applied in
all nodes. In the uncensored sample, the maximum likelihood estimation 
indices are smallest for NAO and SC, meaning that these are
the covariates that mostly influence $\lambda$. The LRT, in
turn, denotes that the combination of the two do not provide significantly
more information than each of these factors by themselves. What is
more, the AIC and the BIC are lowest for the NAO. Therefore, the NAO
is selected as the sole covariate 
for the Poisson-VGAM. Figure \ref{fig:lambda-NAO}
shows that $\lambda$ increases with negative NAO.

NAO, EA, SC (see Figs. \ref{fig:NAO}, \ref{fig:EA},
and \ref{fig:SC}) and their first and second derivatives are also
used as covariates in the NS-CI VGAM to predict the storm-thresholds
and the GPD parameters. The normality and homoscedasticity assumptions
of the VGAM \citep{Rigby05} cannot be rejected for the storm-threshold
and the GPD parameters $x_{0}$ and $\beta$. The incorrelation assumption
is similarly not rejected for the GPD parameters $x_{0}$ and $\beta$, but
should be rejected for the storm-threshold. The latter non-conformity should
be considered when examining the final results.

The statistical model derived from the CMCC-CM (CMCC-A) global circulation-model is, then, compared to the eighteen other models, in the
Supplementary material, Figs. 11 to 18 
show the similarity of CMCC-CM results to other global circulation-models. 
For nodes 7 through 23, the distance between each pair of climate-index
models is relatively short for most cases, except MIROC-ESM-CHEM (MIR-B) and MIROC5 (MIR-C).
The Aitchison and the Kullback-Leibler
distances between $vec_{obs}$ and $vec_{model}$ are shown on Table
\ref{tab:validation}.
The location-parameters of the GPD are presented in Figs. \ref{fig:loc-par-I} and \ref{fig:loc-par-II}.
$\tau$ from
the NS-CI HAC-structures are presented in Figs. \ref{fig:tau-CI-I}
a \ref{fig:tau-CI-II}.

\section{Discussion\label{sec:Discussion}}

\subsection{Pre-analysis (stationarity assumption)\label{subsec:discussion-Pre-analysis}}

The decrease in $E$ and $D$ denote loss of energy and duration of storms 
in future climates. $D$ presents more drastic temporal changes in the northern Catalan Coast.
The $ACC2$ increase in the FF, faster than in the PRNF, suggesting that 
storm-components will present a larger variance over time. 
$ACC2_{E}$ does not behave like $ACC2_{D}$. Possibly, $H_{p}$ has a certain role in lowering the variance of $E$.
The northward decrease in variance of $T_{p}$, observed in Figs. 2 and 3 
of the Supplementary material, 
was also reported for SIMAR hindcasts, in \citet{LinYe16}.
This phenomenon occurs when $T_{p}$ depends heavily on fetch and origin,
rather than being a function of wind pulse characteristics.

As for $E_{year}$,  $\overline H_{s,year}$,  $\overline T_{p,year}$
and $D_{year}$ (see Supplementary material, Figs. 4 to 7), 
$E_{year}$ and $\overline H_{s,year}$ fluctuate from PRNF on, whereas
they have been considerably stationary in PT (see Supplementary material, Fig. 4 and 5). 
The general trend in  $E_{year}$ is a high in the first
quarter of the XXI${}^{\mathrm{st}}$ century, followed by approximately $25\unit{years}$ of low $E_{year}$,
and another quarter of century of high $E_{year}$. $\overline H_{s,year}$ has a cyclicity of 
approximately $50\unit{years}$. $\overline T_{p,year}$ has the same cyclicity as 
 $\overline H_{s,year}$, but it presents stationarity in the PRNF, instead of 
 presenting it in the PT. The time derivatives,
$dE_{year}/dt$, $d\overline H_{s,year}/dt$, $d\overline T_{p,year}/dt$, 
$dD_{year}/dt$ fluctuate periodically, but no clear cycles are detectable (not shown here). 
The reasons behind the clusterings of $E_{year}$,  $\overline H_{s,year}$, 
$\overline T_{p,year}$ and $D_{year}$ peaks
need further atmospheric analysis (see Sub-section \ref{subsec:Discussion-non-stationary-model}),
but the consequences can be outlined. 

$D_{year}$, behaves similarly to $E_{year}$. $E_{year}$ becomes less stable from PRNF
onward. $D_{year}$ and $E_{year}$ behave similarly, due to the definition
of $E$, which includes $D$. 
The low $D_{year}$ and the high $E_{year}$
at the Ebre-Delta in the midst of the XXI${}^{\mathrm{st}}$ century may lead to more
sediment mobility and a loss of resilience of the area, which is already
highly erosive \citep{CIIRC10}. The fact that $E_{year}$ depends
more on a summation of small storms than a great one elevates the
importance of the smaller storms with $1$ to $5\unit{years}$ of
return period. Low lifetime solutions such as Transient Defence Measures
\citep{ASAstot16} would be a plausible solution for these periods.
What can be expected is that these two seasonal features are not going
to be as predictable in the PRNF and FF as in PT, but there are
some remarkable periods in the second half of the XXI${}^{\mathrm{st}}$ century, when
extreme events are present. From the fluctuations of $E_{year}$, 
$\overline H_{s,year}$,  $\overline T_{p,year}$ and $D_{year}$, it can be
perceived that a non-stationary approximation is needed.

\subsection{Stationary model }

The fact that the HAC-structure in Fig. \ref{fig:HAC-tree} is predominant in the AR5-projections
might be due to $H_{p}$ being more dependent of $E$-$D$
in these AR5 projections than in the SIMAR hindcasts \citep{LinYe16}. This means a remarkable
difference between AR5 and SIMAR data. Apparently, the AR5
waves have a lower variability on $H_{p}$ than the SIMAR data, thus leading to this phenomenon.
$E$ and $D$ are averaged values, and a higher correlation can be expected
with data that have lower variability values. In other words, SIMAR
data might be more heteroschedastic than AR5 data, and this affects
the copula definition. Here, the goodness-of-fit of the Gumbel-type HAC
with a ``mean''-type aggregation-method should be acceptable (see Supplementary material, Figs. 8 to 10).

The dependence of $H_{p}$ with the subset $E$-$D$ increases
southward due to the proximity of node 1 to the coast (see Fig. \ref{fig:Study-area}).
The fact that $H_{p}$, $E$ and $D$ have milder values in south-Barcelona
and in Tarragona (not shown here), indicate that storms in the south are less 
energetic and durable than at northern locations. Also, $E$ and $D$
is the strongest related components in all storms, so the more
energy a storm has, the more time it needs to be dissipated, as expected. 

$T_{p}$ becomes independent from the rest of the variables ($E$,
$H_{p}$ and $D$) in the FF. It is observed that, at nodes 1 and
2, $E$, $H_{p}$ and $D$ decrease in the second half
of the XXI${}^{\mathrm{st}}$ century. However, the time series of $T_{p}$ does not present
any trend. Also, except $T_{p}$, the rest of the variables consistently
depend on $D$; as $D$ decreases in the second half
of the XXI${}^{\mathrm{st}}$ century, the other variables behave in the same manner. 
The values of $H_{p}$, $D$ and $E$ are closely inter-connected.
$T_{p}$, on the other hand, is fetch limited, and can hardly surpass
$12\mathrm{s}$, as the most frequent wave direction is related to
a fetch of $550\mathrm{km}$ \citep{G93,ASA08}, several orders of
magnitude lower than Atlantic coasts. The limitation by fetch
can also be observed on the $H_{p}$ data, for all time-frames.
The temporal and spatial variability of $H_{p}$ are greater, however,
than those of $T_{p}$. The main storm impact is thus reduced to
isolated energetic events, with no previous warning nor further replicas.
The isolated nature of such events will make storm forecasting a fundamental
management tool in the future, based on causal factors, rather than warning signals of the surrounding environment.

\subsection{Non-stationary model \label{subsec:Discussion-non-stationary-model}}

The  storm-thresholds of the non-stationary model, in all the nodes, fall on the linear part
of the excess-over-treshold graphs for PT, PRNF, and FF (see Fig.
\ref{fig:excess-over-threshold}). Therefore, these thresholds are
defining extreme events \citep{Tolosana-Delgado2010}.

According to Fig. \ref{fig:lambda-NAO}, $\lambda$
increases with negative NAO. This contradicts \citet{Nissen2014},
who stated that positive NAO are more favourable for cyclone intensification, opposite to the findings here. Hence, further research is needed to help revise the relationship between $\lambda$ and NAO, and since NAO is
strongly related to temperature changes, 
Climate-Change indirectly affects storminess at the Catalan
Coast.

In the censorship analysis at nodes 7 and 21, cases with on-shore and 
 off-shore winds have presented better metrics that the general model herein presented. 
When the model is
built with the whole storm sample, the interaction of the covariates
leads to more variability among the global circulation-models. This analysis has also reinforced
the initial hypothesis that onshore winds are correlated with NAO
and offshore winds with SC, which is plausible for the study area.  Regarding the uncensored
sample, the most influencing covariates for storm-threshold 
are: $\mathrm{NAO}$, $d\mathrm{^{2}EA}$,
and $\mathrm{SC}$. The covariates mostly affecting the GPD location
parameter $x_{0}$ of each storm-intensity variable are: $d\mathrm{SC}$
for the $E$; $\mathrm{SC}$ for $H$ and $T_{p}$; and $\mathrm{EA}$,
for $D$. The most influencing factors on the GPD scale-parameter
$\beta$ of each storm-intensity variable are: $d\mathrm{^{2}EA}$ for the
$E$; $d^{2}\mathrm{EA}$ and $d^{2}\mathrm{SC}$ for $H$; $\mathrm{NAO}$
for $T_{p}$, and $d\mathrm{SC}$ for $D$. 
From all the possible
combinations with climate-indices and their time derivatives,
the abovementioned covariates have been the ones that presented minimum
AIC and BIC, plus lower p-values of LRT. The suitability of these covariates strongly suggests that storms
are more affected by the dynamics (sea level pressure
gradients) of climate-indices than the climate-indices themselves.
In other words, gradients in atmospheric change can 
lead to an outcome different from that of regular shifts of atmospheric states. 

Regarding the $99^{\mathrm{th}}$ quantile in Figs. 11 to 18 of the Supplementary material, 
both amplitude, phase and trend
of the signals present similar patterns in all global circulation-models, although the oscillations
do not necessarily coincide among themselves (summarized in Figs. 11 to 18 of the Supplementary material).
Stronger disagreement at nodes 1 and 5 can also be understood, because of the strong bimodality that exists on the southern
part of the Catalan Coast \citep{G93,Grifoll16b}. The Aitchison
and Kullback-Leibler distances between $vec_{obs}$ and $vec_{model}$
\ref{tab:validation} are of order $1$, which is the order of magnitude of the module of the vectors, 
in all the validating nodes. This indicates that the proposed model has been well validated. 

The obtained results do not indicate that
Climate-Change is the main contributor to the switch in storm-patterns. It is not certain
to what extent this is related to natural variability
of large scale indices and how it is affected by the anthropogenic
footprint \citep{Trenberth15}. Such an explanatory analysis denotes
that in this time period, the CMCC-CM global circulation-model presents a climate in which
the superposition of both natural variability and greenhouse gases
will lead to this change. Regardless of each component's contribution,
this information can be useful to tackle problematic
seasons in the future. 

The trends of the GPD location-parameters of storm-intensity variables
(see Figs. \ref{fig:loc-par-I} and \ref{fig:loc-par-II})
determine their general behaviour.
So that where the location-parameters
of $E$, $H_{p}$ and $T_{p}$ decrease in time, there should also be a linear decrease of the variables. 
There is much noise for all variables except $T_{p}$.
 The trends of the GPD location-parameters
$x_{0}$ of $E$, $H_{p}$, and $T_{p}$ are either constant or downward.
$D$ clearly increases in time
at the northern Catalan Coast. This increase may have a relevant impact on harbours,
which would require adaptive engineering to face switches in storm-wave patterns
 and sea-level-rise \citep{Burcharth2014,ASA2016a}. Meanwhile, the trend of $D$ is negative at the southern Catalan Coast.
The decrease in $E$ has been suggested in Subsection \ref{subsec:discussion-Pre-analysis},
but the increase in $D$ at the northern Catalan Coast is a new information that has only been clarified by the non-stationary model. 

 As for the semi-dependence among storm-components, $\tau$ (see Figs.
\ref{fig:tau-CI-I} to \ref{fig:tau-CI-II})
 values are more constant at the north coast than near the Ebre Delta (south coast), where water
depths are shallower. That is to say that, wave conditions present
more variability in shallower waters. $\tau_{\left(E,D\right)}$ has
a considerable upward trend at all nodes. This might be explained
by a decreasing role of wave-height, and a predominant
role of $D$ as the local storm feature. There also seems to be a cyclical
variation in dependence among variables, whose cause
should be explored in future work. It can also be noted that the
peak of $\tau_{\left(\left(E,D\right),H\right)}$ in the 
period 2000-2050 shows a particular dependence of $H_{p}$ with
respect to $D$, hinting a concurrence of extreme conditions
for wave-height and storm-duration.

\section{Conclusions\label{sec:Conclusions}}

The extreme wave-climate under a RCP8.5 Climate-Change scenario has been
characterised for a fetch-limited environment (Catalan Coast). For this purpose, a non-stationary model for
the extreme wave-climate in the period 1950-2100 has been built. The pre-analysis under the stationary
assumption provides a first assessment of the AR5 projected storms.
It suggests that wave-storms might be dependent on time, stressing
the importance of a non-stationary approach. In addition, the stationary
model suggests a HAC-structure for this non-stationary approach. 

The non-stationary model establishes two types of covariates: a) time
and b) climate-indices. The first type indicates the necessity of
a non-stationary approach, whereas b) analyses the effects of
climate-indices, and their first
and second time-derivatives. Storminess appears to depend specially
on NAO, as the negative NAO may be associated with storm intensification. Regarding
storm-thresholds and the parameters of the GPDs, they are most influenced
by the dynamics of climate-indices, rather than by the value of
the indices. Location-parameters decrease with time for all 
variables, except for storm duration ($D$) at the northern part
of the Catalan Coast. HAC dependence-parameters ($\tau$)
between storm energy ($E$) and duration ($D$) present a considerable upward
trend in time. Also, the peak of $\tau_{\left(\left(E,D\right),H\right)}$
in the period 2000-2050 can be translated as a 
climatic co-existence (under present conditions) of extreme conditions for wave-height ($H_{p}$) and
storm duration, $D$.

\paragraph*{Funding}

This paper has been supported by the European project CEASELESS
(H2020-730030-CEASELESS), the Spanish national projects PLAN-WAVE
(CTM2013-45141-R) and the MINECO FEDER Funds co-funding CODA-RETOS
(MTM2015-65016-C2-2-R). As a fellow group, we would also like to thank
the Secretary of Universities and Research of the department of Economics
of the Generalitat de Catalunya (Ref. 2014SGR1253, 2014SGR551). The second
author acknowledges the Ph.D. scholarship from the Generalitat de Catalunya
(DGR FI-AGAUR-14). 

\paragraph*{Acknowledgements}

The support of the Puertos del Estado, in providing the buoy data and the SIMAR model outputs, is also duly
appreciated.

\bibliographystyle{klunamed}
\addcontentsline{toc}{section}{\refname}\bibliography{Seasons,references_sed_transport_models,NHMlib,Copula}


\noindent \begin{flushleft}
\begin{figure}[H]
\includegraphics[bb=100bp 300bp 704bp 704bp,clip]{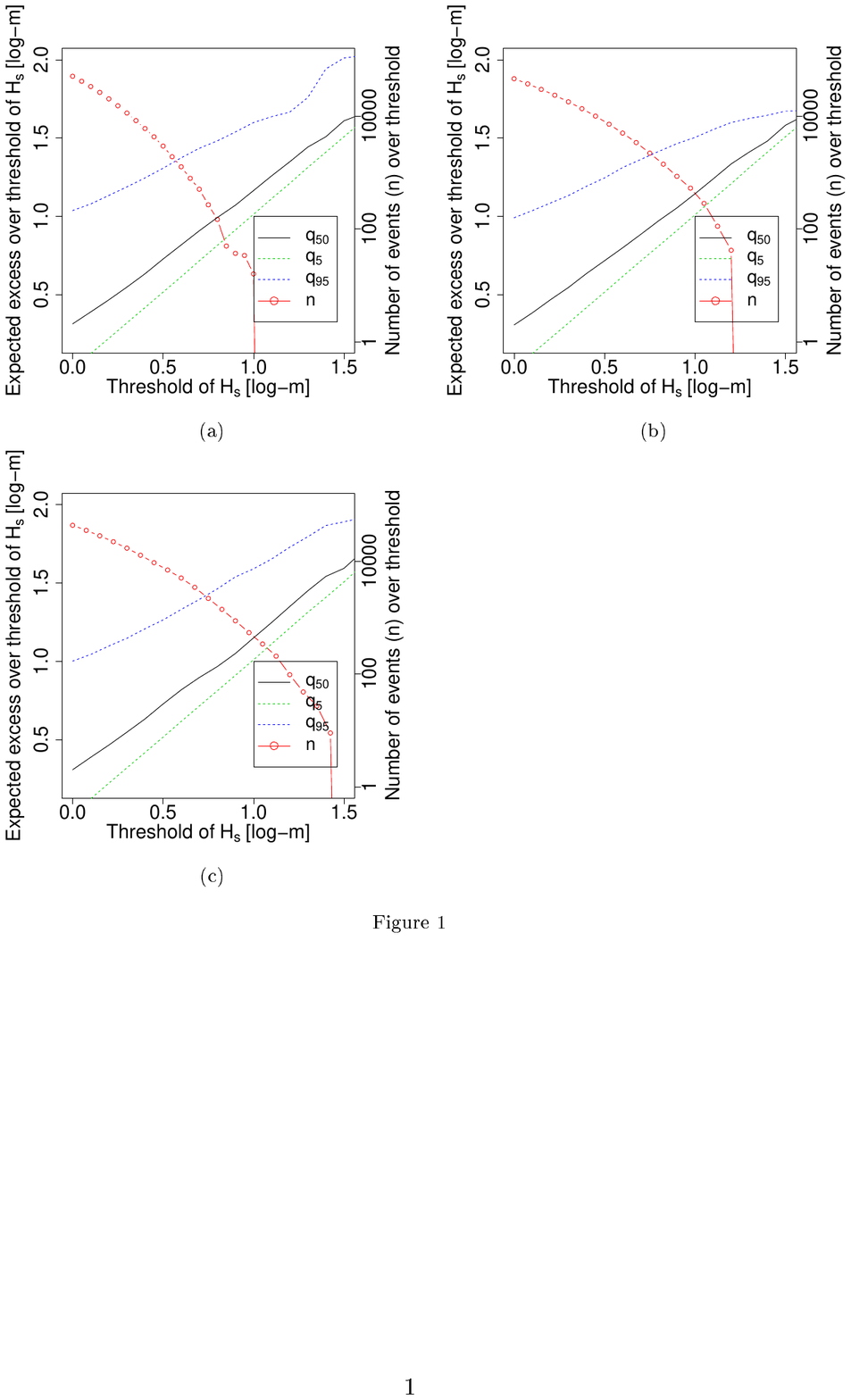}
\caption{Excess-over-threshold plots at node 12, in a) past (PT), b) present-near-future (PRNF), and c)
far-future (FF) time frames. The red line denotes the number of events $\left(n\right)$
over the threshold. \label{fig:excess-over-threshold}}
\end{figure}

\par\end{flushleft}

\begin{figure}[H]
\begin{centering}
\includegraphics[bb=60bp 150bp 480bp 504bp,clip,width=7cm]{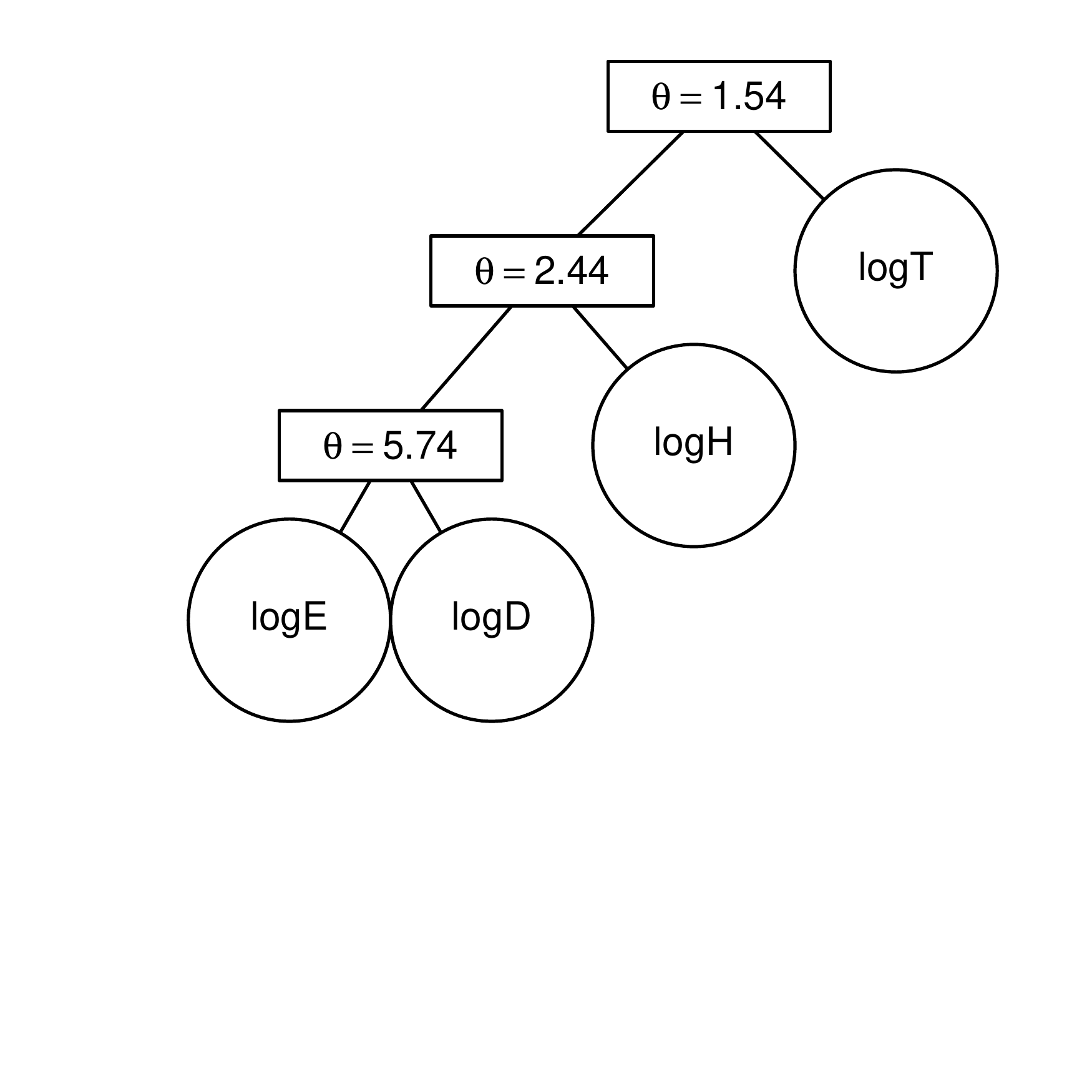}
\par\end{centering}
\caption{Example of HAC-structure, at node 12, in past (PT). The circles enclose the analysed storm variables, and the $\theta$ is the HAC-dependence-parameter.\label{fig:HAC-tree}}
\end{figure}
\begin{figure}[H]
\includegraphics[bb=0bp 0bp 504bp 504bp,clip,scale=0.6]{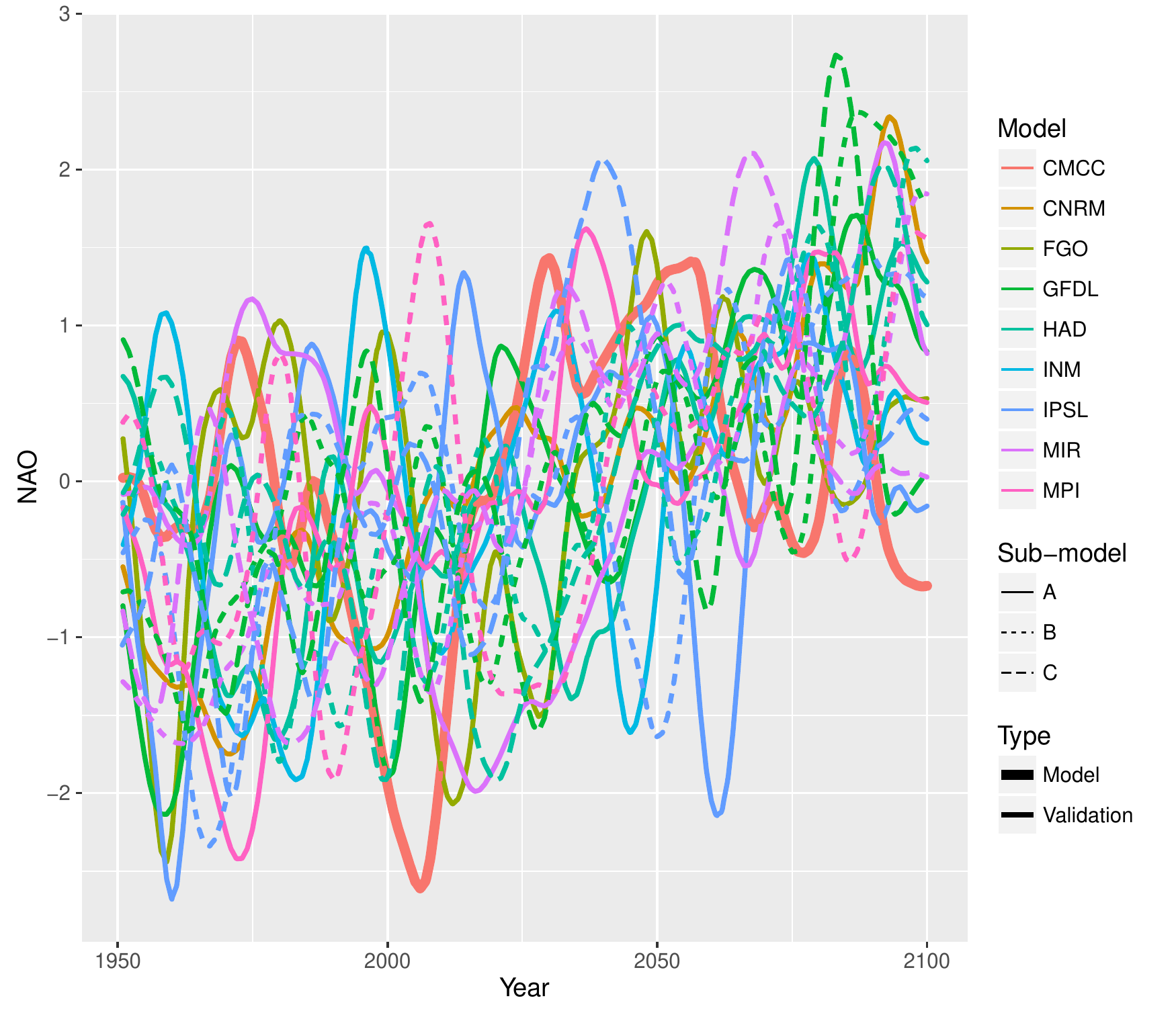}
\caption{Temporal evolution of NAO index from the global circulation-model monthly outputs
(see Table \ref{tab:GCM_models}). NAO is represented by an adimensional index, scaled to have a mean
value equal to zero and a variance equal to unity.\label{fig:NAO}}
\end{figure}


\begin{figure}[H]
\includegraphics[bb=0bp 0bp 504bp 504bp,clip,scale=0.6]{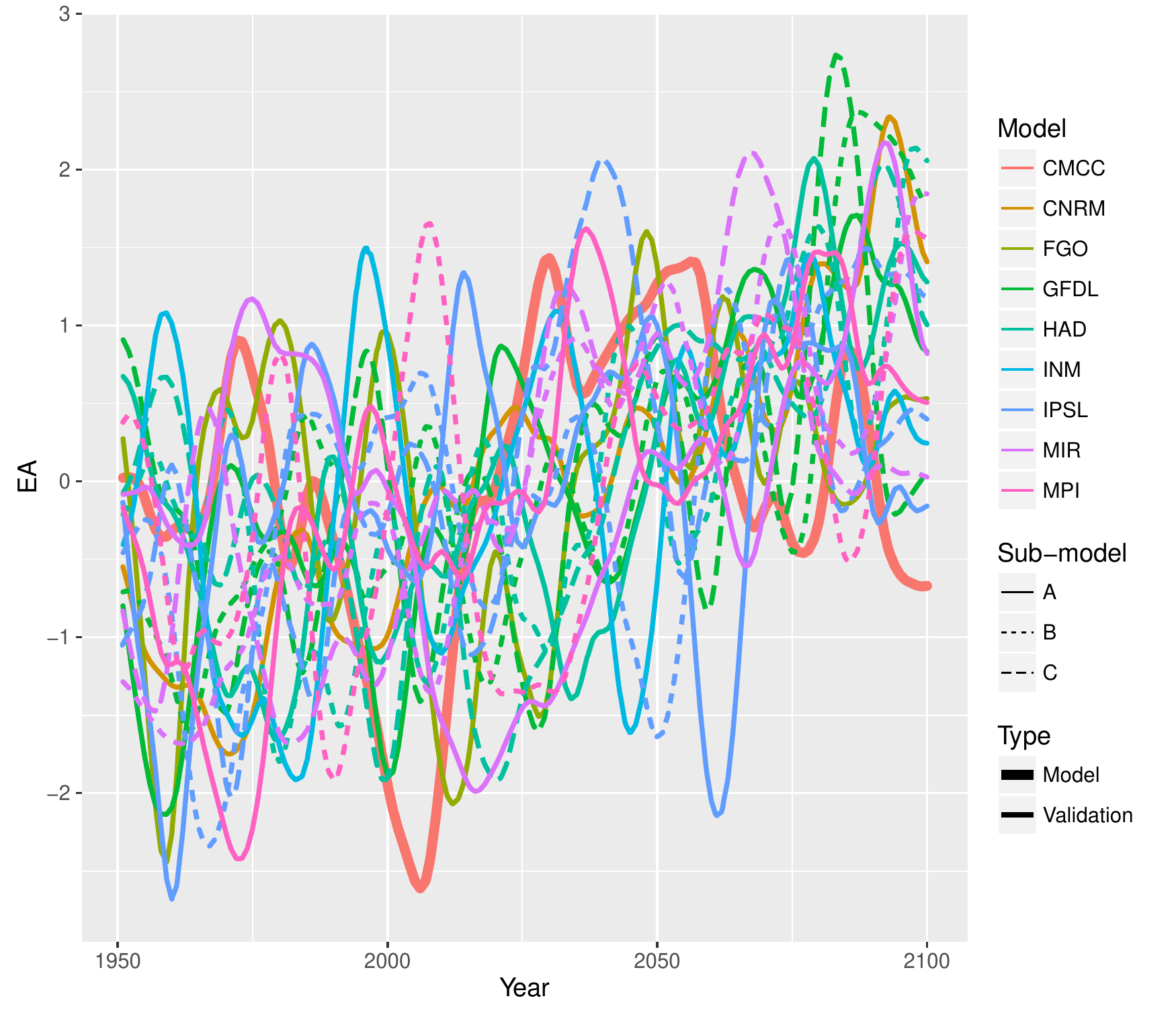}
\caption{Temporal evolution of EA index from the global circulation-model monthly outputs
(see Table \ref{tab:GCM_models}). EA is represented by an adimensional index.\label{fig:EA}}
\end{figure}
\begin{figure}[H]
\includegraphics[bb=0bp 0bp 504bp 504bp,clip,scale=0.6]{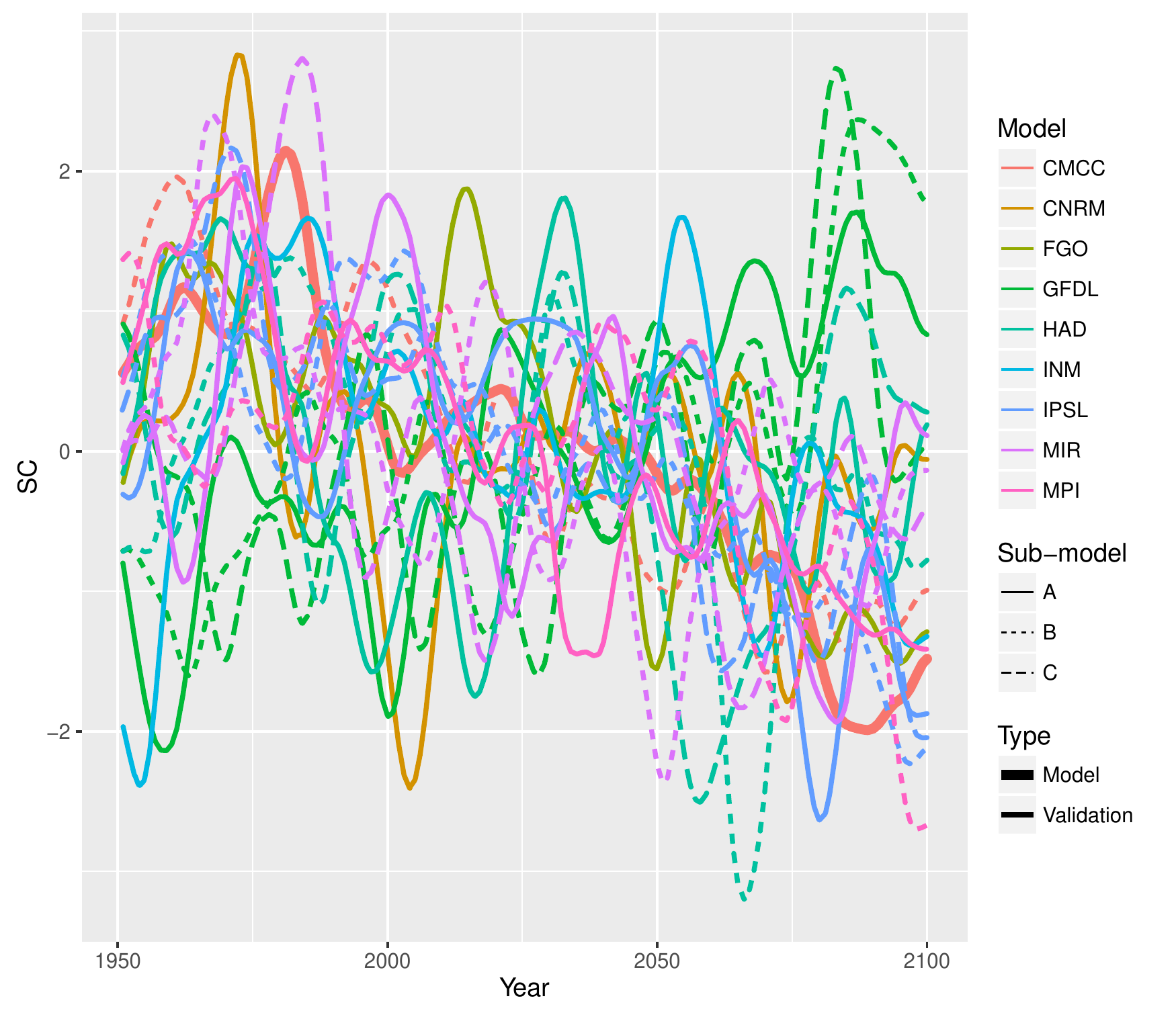}
\caption{Temporal evolution of SC index from the global circulation-model monthly outputs (see Table \ref{tab:GCM_models}).
SC is represented by an adimensional index.
\label{fig:SC}}
\end{figure}

\begin{figure}[H]
\includegraphics[bb=0bp 0bp 504bp 504bp,clip,scale=0.6]{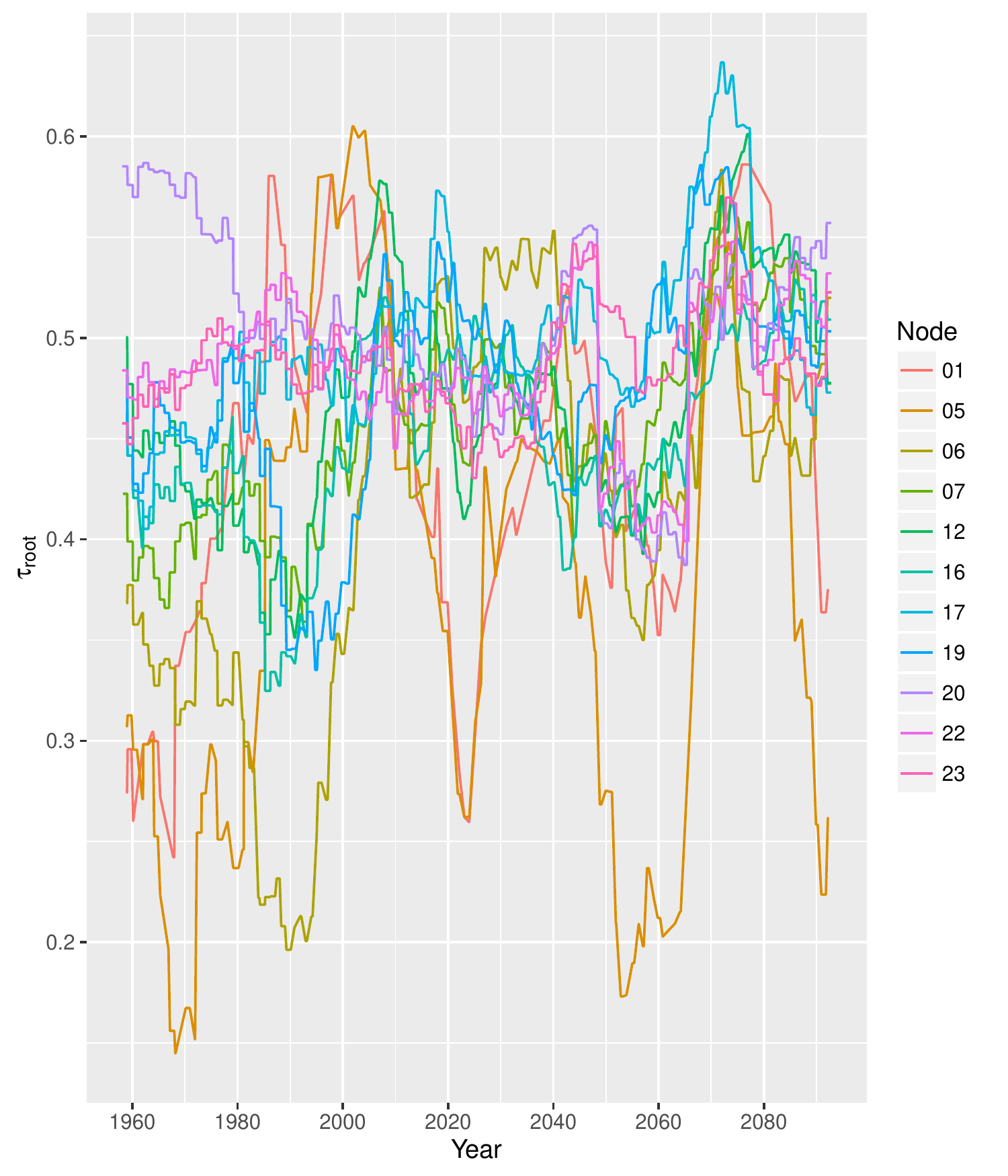}
\caption{Non-stationary $\tau_{root}$ dependence parameter \citep{Kendall37} at the root 
nesting level of the HAC structure. The marginal distributions are
fitted with the VGAM, with time as the sole covariate (NS-T).
The colours represent different nodes. \label{fig:tau-time-root}}
\end{figure}
\begin{figure}[H]
\includegraphics[bb=0bp 0bp 504bp 504bp,clip,scale=0.6]{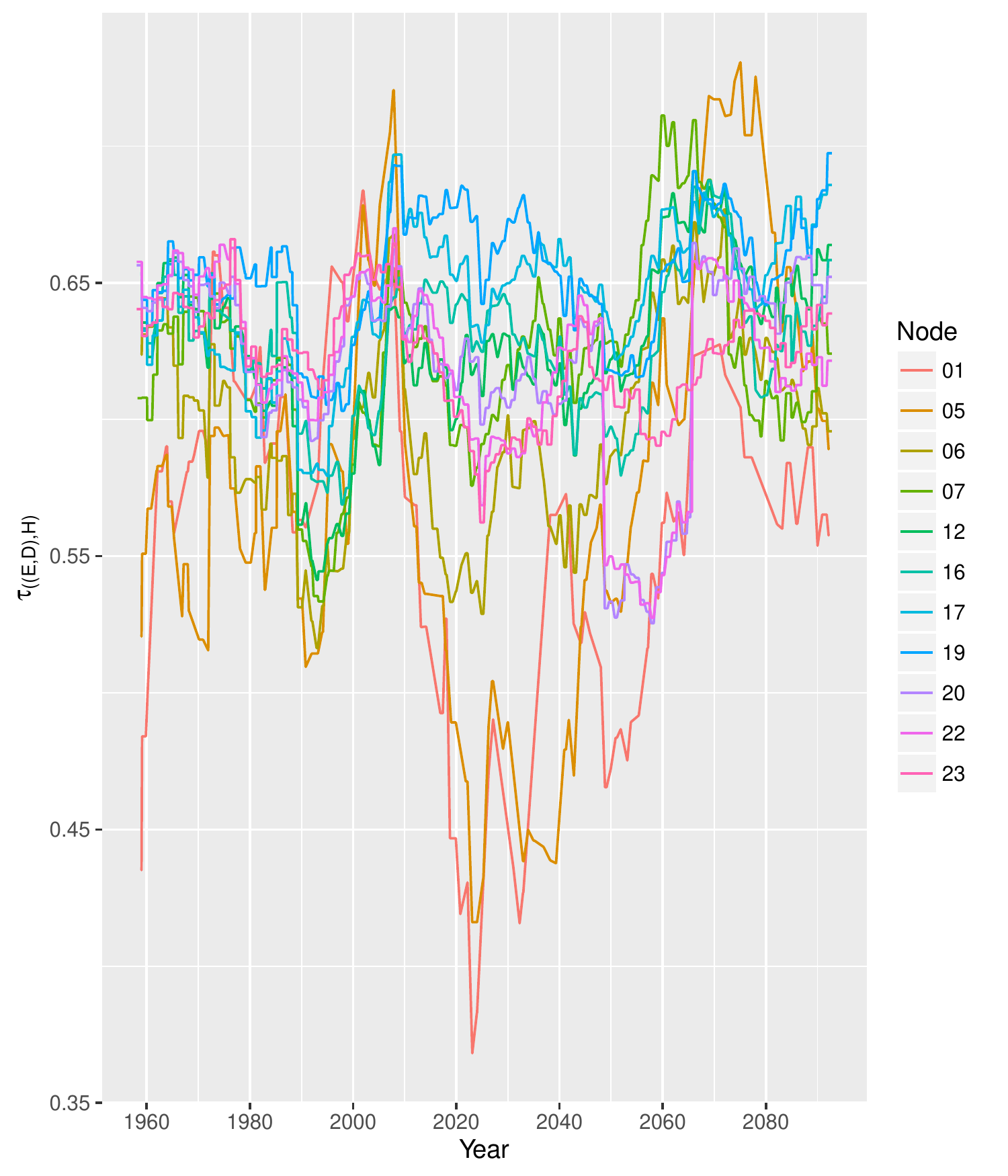}
\caption{Non-stationary $\tau_{\left(\left(E,D\right),H\right)}$
dependence parameter at the$\left(\left(E,D\right),H\right)$
nesting level of the HAC structure. 
The marginal distributions are
fitted with the VGAM, with time as the sole covariate (NS-T).\label{fig:tau-time-EDH}}
\end{figure}
\begin{figure}[H]
\includegraphics[bb=0bp 0bp 504bp 504bp,clip,scale=0.6]{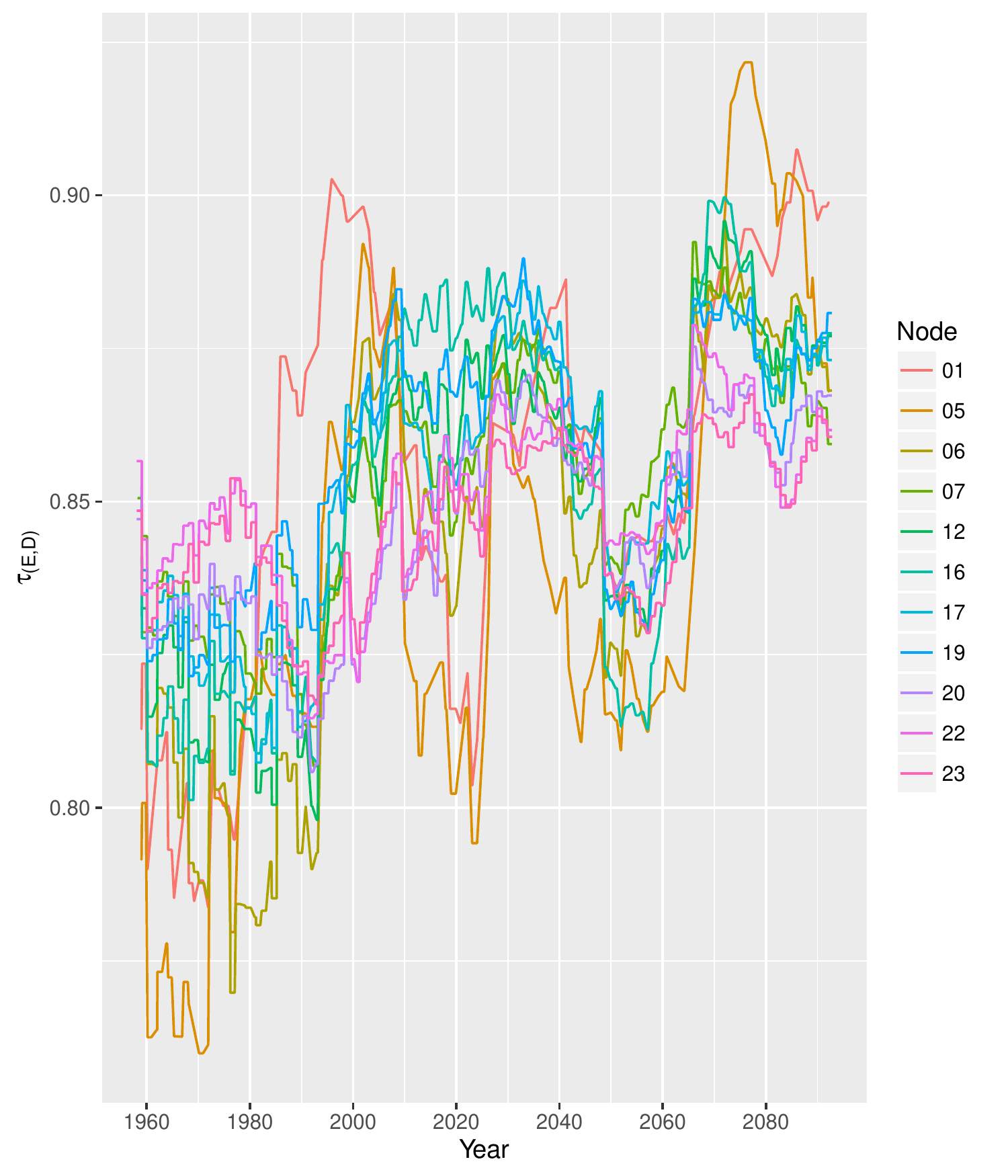}
\caption{Non-stationary $\tau_{((E,D))}$ dependence parameter at the (E,D) nesting level of the HAC.
The marginal distributions are
fitted with the VGAM, with time as the sole covariate (NS-T).
 \label{fig:tau-time-ED}}
\end{figure}

\begin{figure}[H]
\begin{raggedright}
\begin{minipage}[t]{0.75\columnwidth}%
\begin{center}
\includegraphics[width=13cm]{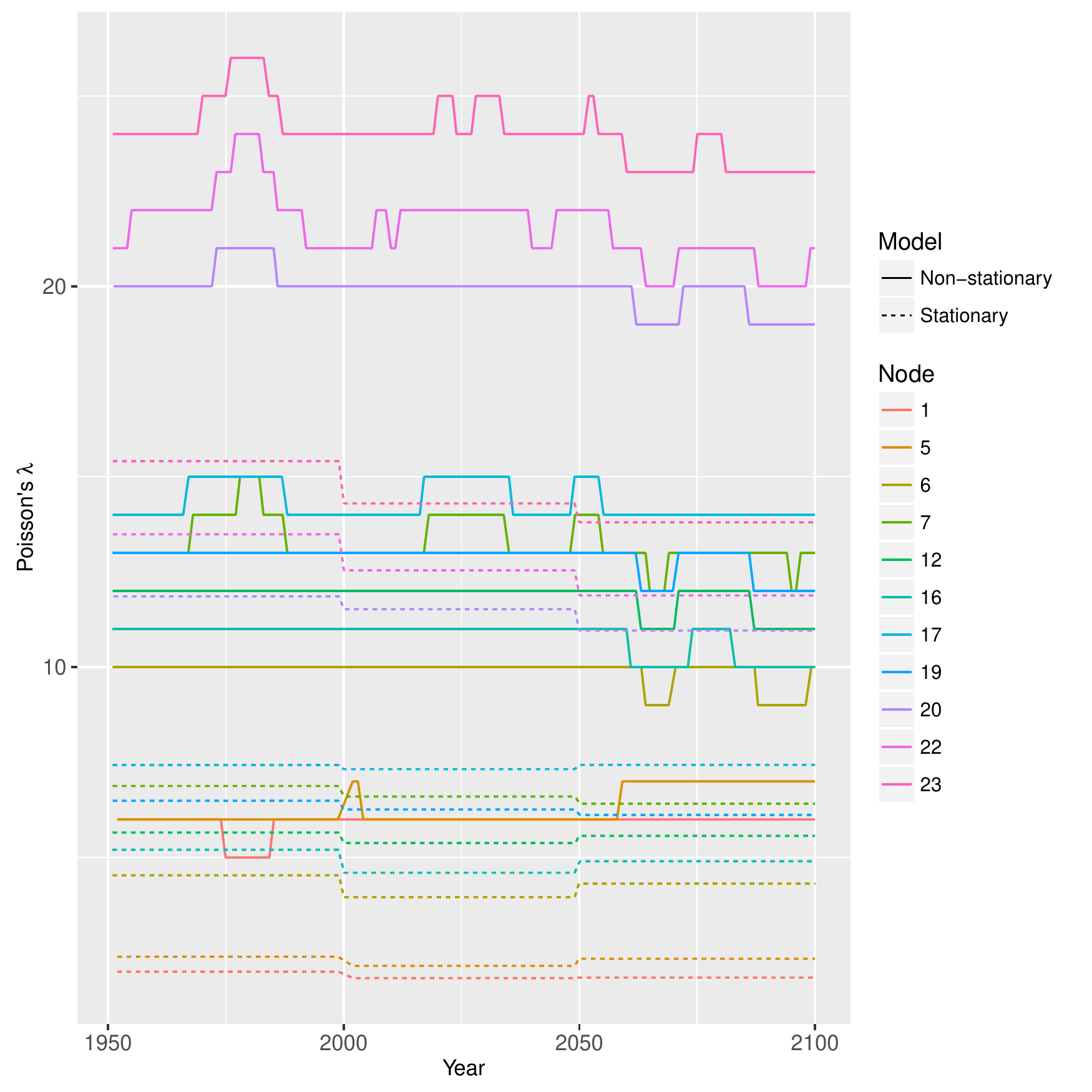}
\par\end{center}%
\end{minipage}
\par\end{raggedright}

\caption{Storminess-index function $(\lambda)$ for the stationary and non-stationary
models, the latter using time as covariate (NS-T).
\label{fig:lambda-time}}
\end{figure}


\begin{figure}[H]
\begin{raggedright}
\begin{minipage}[t]{0.75\columnwidth}%
\begin{center}
\includegraphics[width=13cm]{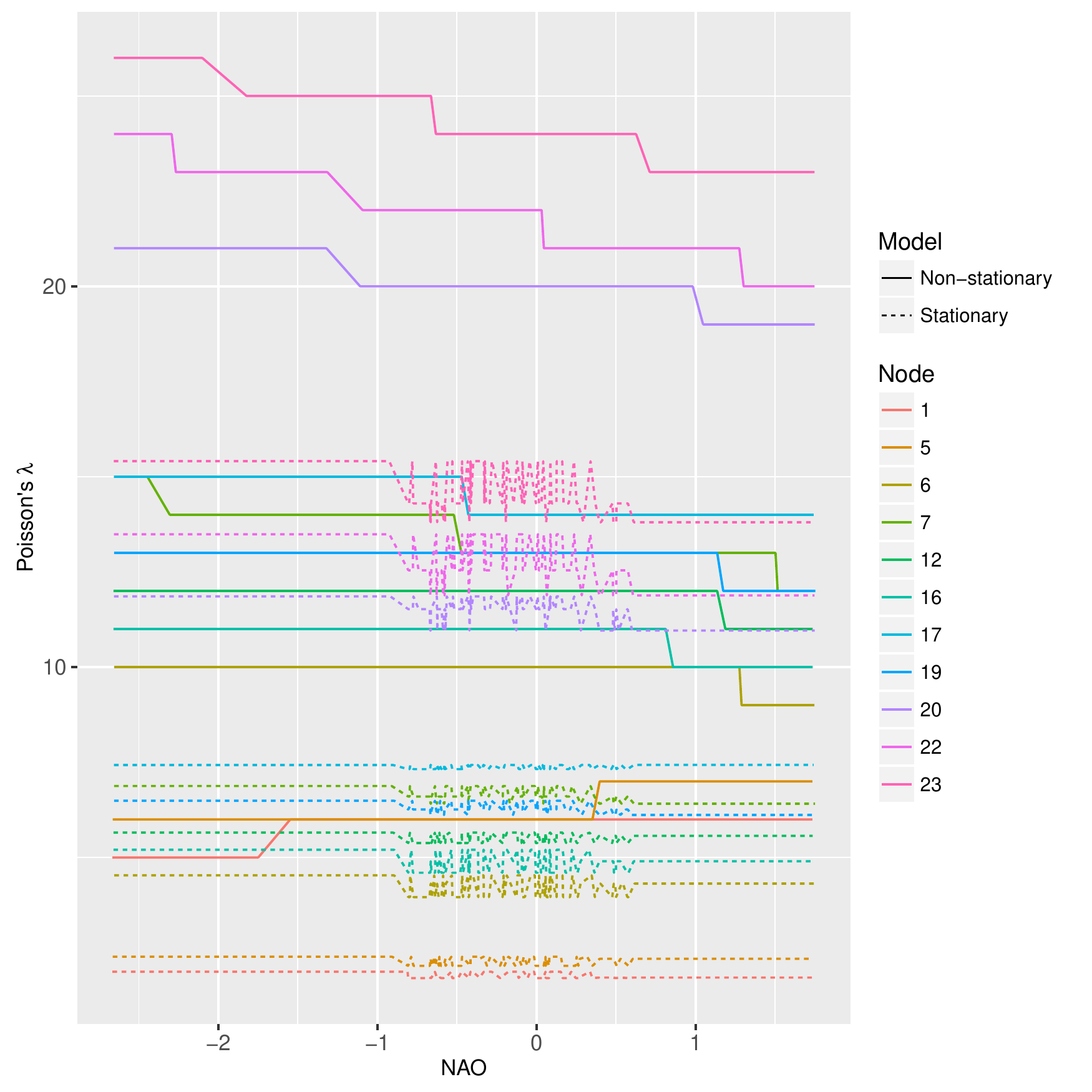}
\par\end{center}%
\end{minipage}
\par\end{raggedright}

\caption{Storminess-index function $(\lambda)$ for stationary
and non-stationary models, the latter using NAO
as covariate (from the CMCC-CM, or CMCC-A, model, NS-CI). \label{fig:lambda-NAO}}
\end{figure}

\begin{sidewaysfigure}
\includegraphics[bb=0bp 120bp 430bp 700bp,clip,angle=270]{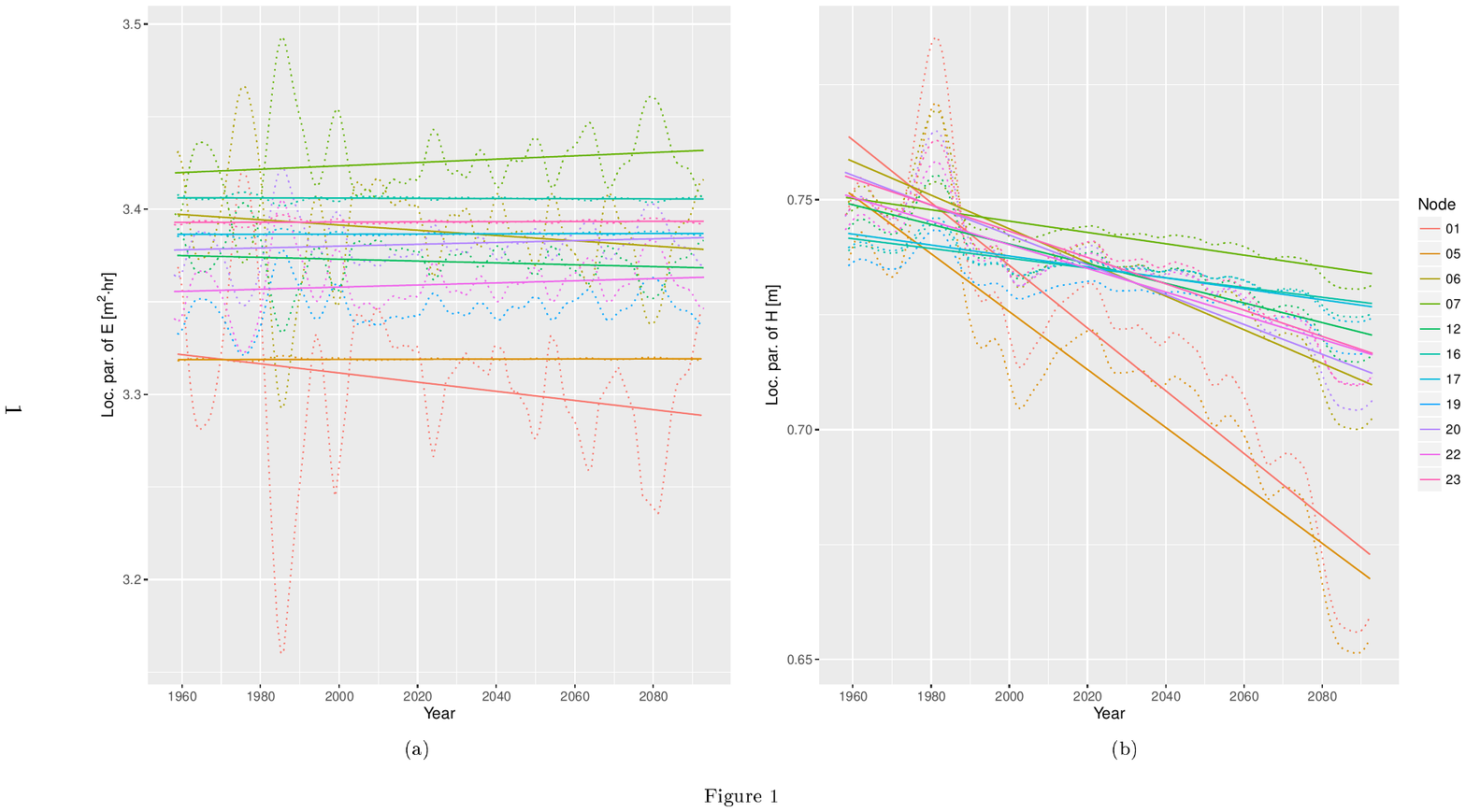}
\caption{Non-stationary GPD location-functions $\left(x_{0}\right)$ for a)
wave energy $\left(E\right)$ and b) significant wave-height at the peak $\left(H_{p}\right)$ using
VGAM (GPD distribution) with climate-indices as covariates:
$E\sim\left(GPD\left(\mu\left(dSC\right),\sigma\left(d^{2}EA\right),\xi\right)\right)$
and $H_{p}\sim\left(GPD\left(\mu\left(SC\right),\sigma\left(d^{2}EA,d^{2}SC\right),\xi\right)\right)$.
The discontinuous lines show the time variation of the location-parameter and the solid lines represent their
linear trend. The colours represent
different nodes (see Fig. \ref{fig:Study-area}).
\label{fig:loc-par-I}}
\end{sidewaysfigure}


\begin{sidewaysfigure}
\includegraphics[bb=0bp 120bp 430bp 700bp,clip,angle=270]{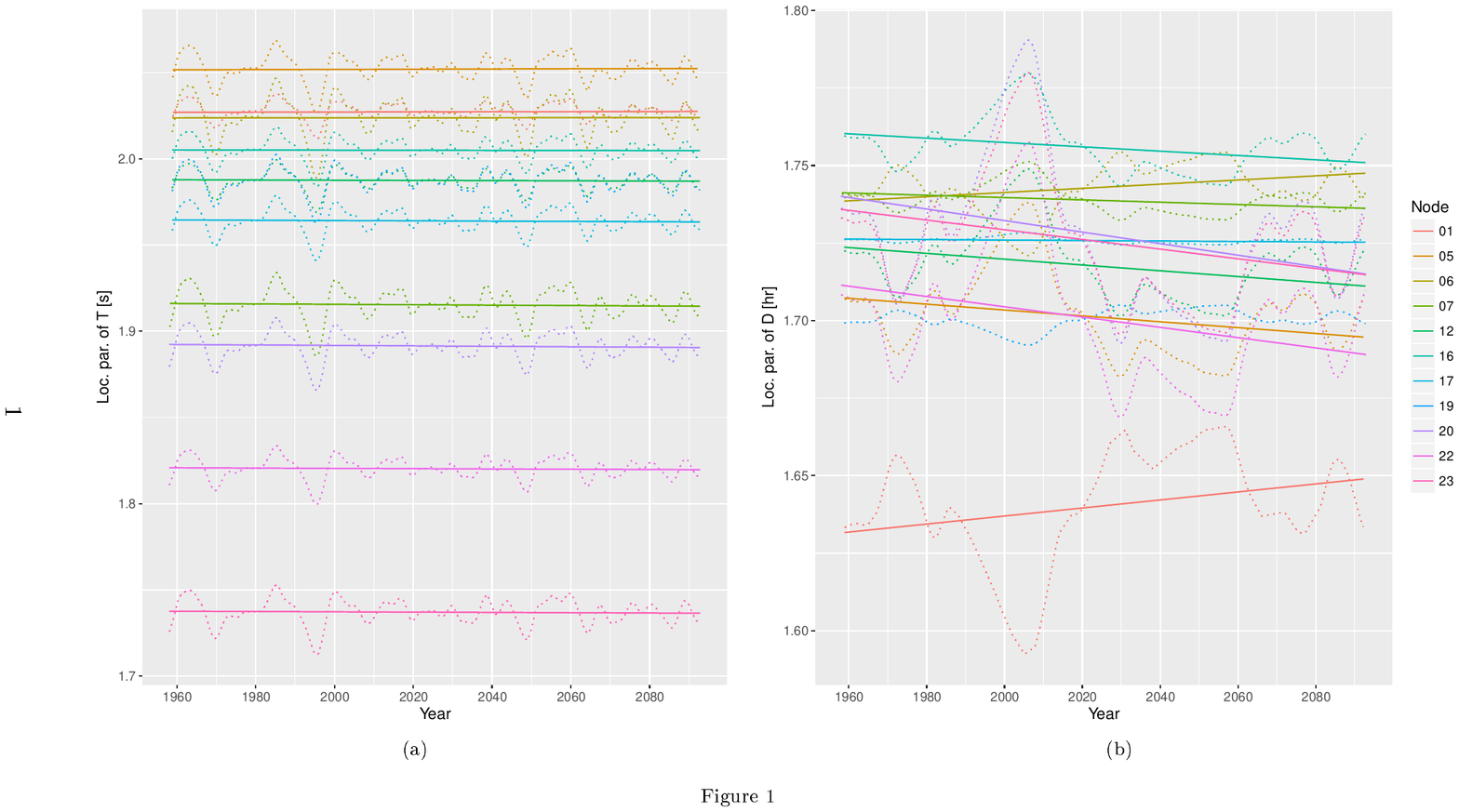}
\caption{Non-stationary GPD location-parameters $\left(x_{0}\right)$ for a)
peak-period $\left(T_{p}\right)$ and b) storm-duration $\left(D\right)$
using VGAM (GPD distribution) with climate-indices as
covariates: $T_{p}\thicksim\left(GPD\left(x_{0}\left(SC\right),\beta\left(NAO\right),\xi\right)\right)$
and $D\sim\left(GPD\left(x_{0}\left(EA\right),\beta\left(dSC\right),\xi\right)\right)$.
The discontinuous lines show the time variation of the location function
and the solid lines represent their
linear trend. The colours represent
different nodes (see Fig. \ref{fig:Study-area}).
\label{fig:loc-par-II}}
\end{sidewaysfigure}

\begin{sidewaysfigure}
\includegraphics[bb=0bp 120bp 430bp 700bp,clip,angle=270]{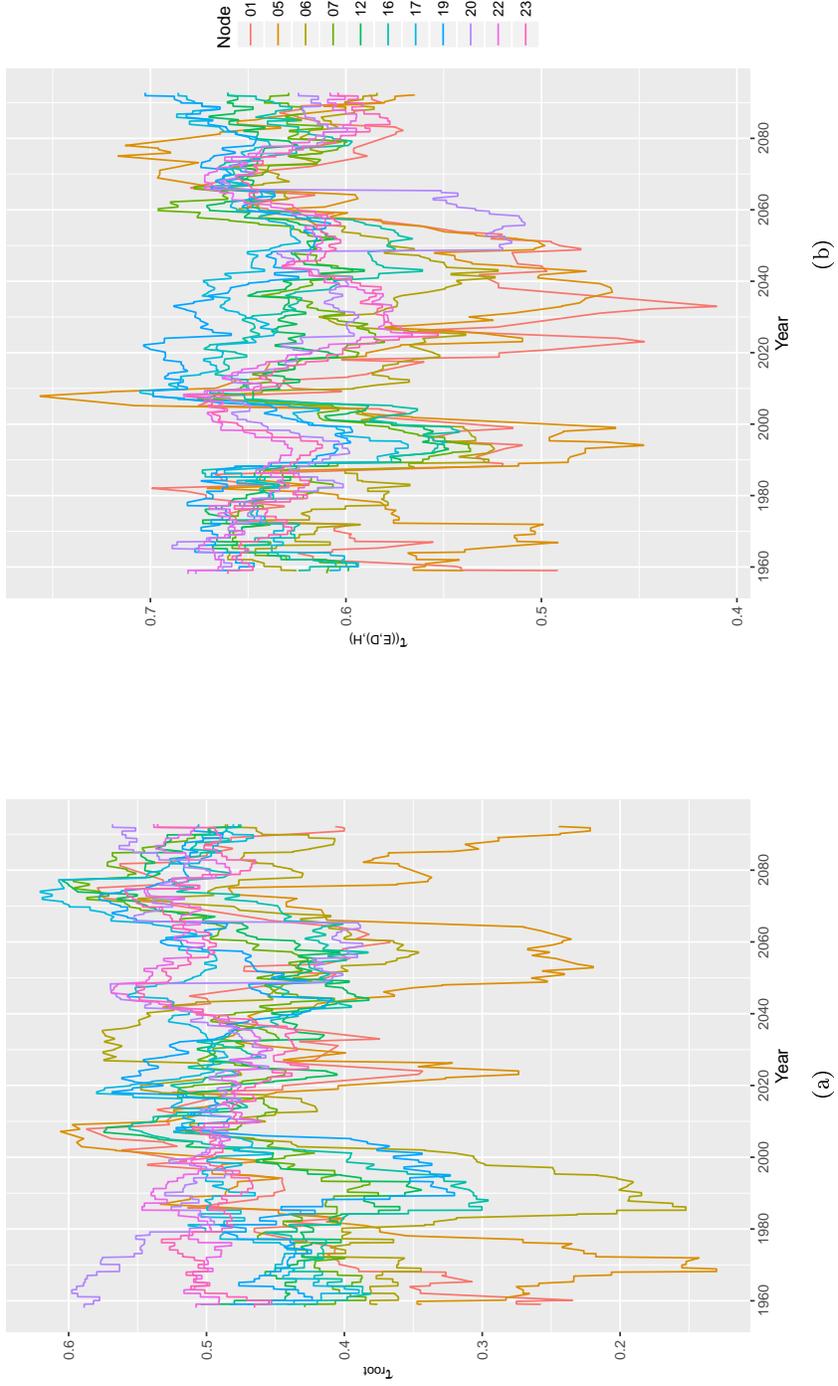}
\caption{Non-stationary $\tau_{root}$ and $\tau_{\left(\left(E,D\right),H\right)}$
dependence-parameter \citep{Kendall37} at the root and $\left(\left(E,D\right),H\right)$
nesting levels of the HAC structure. The marginal distributions are
fitted with the VGAM with climate-indices as covariates
(NS-CI). The colours represent different nodes (see Fig. \ref{fig:Study-area}). \label{fig:tau-CI-I}}
\end{sidewaysfigure}

\begin{figure}[H]
\includegraphics[bb=0bp 0bp 504bp 504bp,clip,scale=0.6]{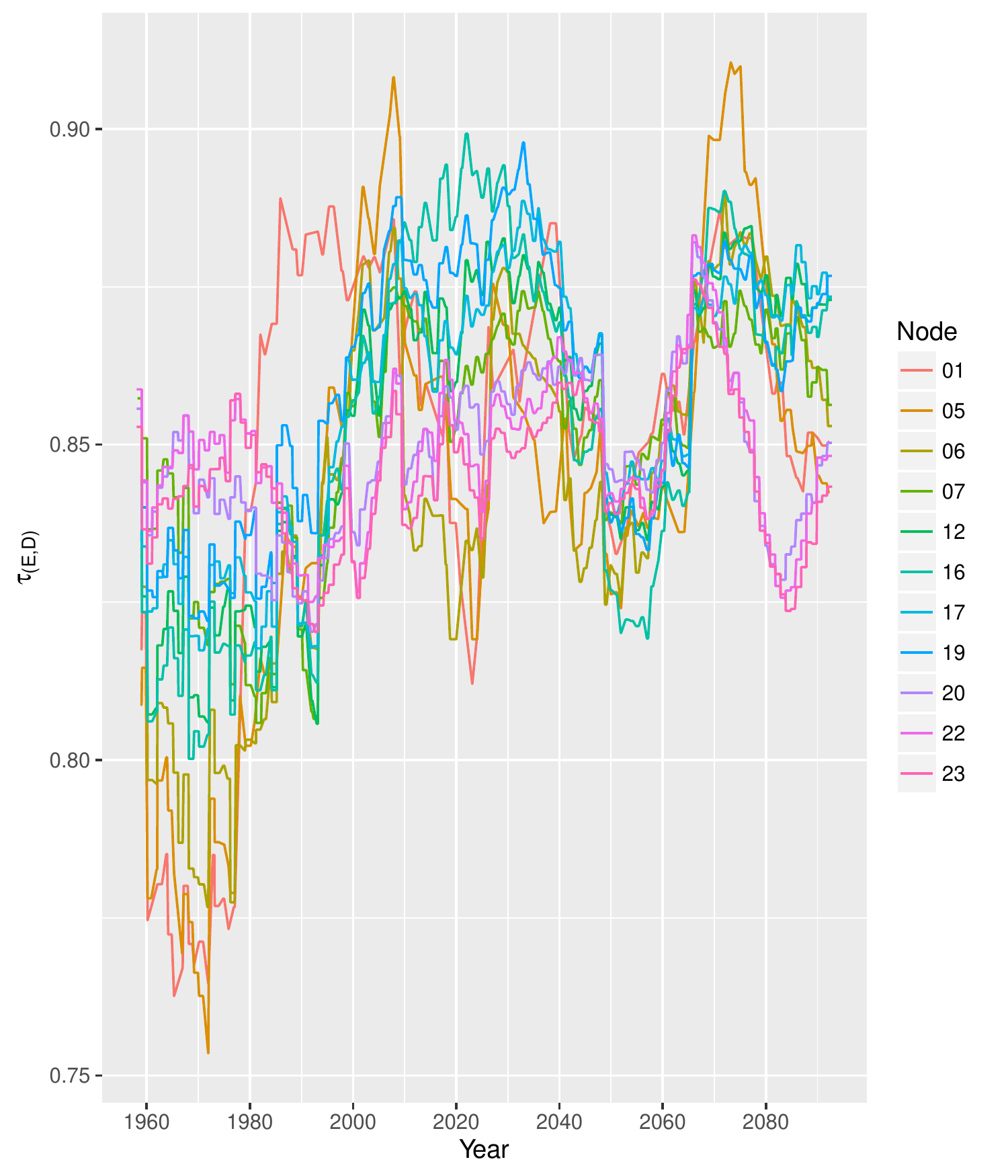}
\caption{Non-stationary $\tau_{((E,D))}$ dependence parameter
at the (E,D) nesting level of the HAC. \label{fig:tau-CI-II}}
\end{figure}
\begin{table}[H]
\caption{\label{tab:GCM_models}Global circulation-models from CMIP5 experiment \citep{Taylor12a}
that are considered in this study. The latitude and longitude columns
denote the grid size.}
\centering{}%
\begin{tabular}{llll}
\hline 
Acronym & Global circulation-model & Latitude & Longitude\tabularnewline
 &  & grid size ($^{\circ}$) & grid size($^{\circ}$)\tabularnewline
\hline 
CMCC\_A & CMCC-CM & 0.7484 & 0.75\tabularnewline
CMCC\_B & CMCC-CMS & 3.7111 & 3.75\tabularnewline
CNRM\_A & CNRM-CM5 & 1.4008 & 1.40625\tabularnewline
FGO\_A & FGOALS-G2 & 2.7906 & 2.8125\tabularnewline
GFDL\_A & GFDL-CM3 & 2 & 2.5\tabularnewline
GFDL\_B & GFDL-ESM2G & 2.0225 & 2\tabularnewline
GFDL\_C & GFDL-ESM2M & 2.0225 & 2.5\tabularnewline
HAD\_A & HadGEM2-AO & 1.25 & 1.875\tabularnewline
HAD\_B & HadGEM2-CC & 1.25 & 1.875\tabularnewline
HAD\_C & HadGEM2-ES & 1.25 & 1.875\tabularnewline
INM\_A & INM-CM4 & 1.5 & 2\tabularnewline
IPSL\_A & IPSL-CM5A-LR & 1.8947 & 3.75\tabularnewline
IPSL\_B & IPSL-CM5B-LR & 1.8947 & 3.75\tabularnewline
IPSL\_C & IPSL-CM5A-MR & 1.2676 & 2.5\tabularnewline
MIR\_A & MIROC-ESM & 2.7906 & 2.8125\tabularnewline
MIR\_B & MIROC-ESM-CHEM & 2.7906 & 2.8125\tabularnewline
MIR\_C & MIROC5 & 1.4008 & 1.40625\tabularnewline
MPI\_A & MPI-ESM-LR & 1.8653 & 1.875\tabularnewline
MPI\_B & MPI-ESM-MR & 1.8653 & 1.875\tabularnewline
\hline 
\end{tabular}
\end{table}
\begin{table}
\caption{Validation of the proposed model by computing the Aitchison and the Kullback-Leibler distances between $vec_{obs}$
and $vec_{model}$ (see eqs. \ref{eq:aitch-obs} and \ref{eq:aitch-mod}).\label{tab:validation}}

\centering{}%
\begin{tabular}{llll}
\hline 
SIMAR/buoy  & AR5  & Ait.dist($vec_{obs},vec_{model}$)& KL.dist($vec_{obs},vec_{model}$)\tabularnewline
node & node &(Aitchison distance)&(Kulback-Leibler distance) \tabularnewline
\hline 
N1  & 23  & 0.52  &  0.07\tabularnewline
N3  & 22  & 0.81  &  0.16\tabularnewline
N4  & 20  & 0.18  &  0.01 \tabularnewline
N7 & 19  & 0.45  &  0.05\tabularnewline
N8  & 17  & 0.54 &  0.07\tabularnewline
C1 &  16  & 0.20 &  0.01\tabularnewline
C3  & 12  & 0.26 &  0.02\tabularnewline
C4  & 07 & 0.26  &  0.02\tabularnewline
C5  & 06 & 0.96  &  0.24\tabularnewline
S4 & 5 & 1.31 &  0.30\tabularnewline
S7 & 1 & 0.98  &  0.23\tabularnewline
PdE-Begur  & 20 & 0.96 &  0.24 \tabularnewline
PdE-BCN-I  & 12  & 1.31  &  0.41\tabularnewline
\hline 
\end{tabular}
\end{table}
\end{document}